\documentclass[aps,prd,twocolumn,superscriptaddress,nofootinbib,longbibliography,showkeys,colorlinks=true,citecolor=blue,linkcolor=blue,urlcolor=blue]{revtex4-1}

\usepackage[T1]{fontenc}
\usepackage[utf8]{inputenc}
\usepackage{amsmath,amssymb,bm}
\usepackage{graphicx}
\usepackage{orcidlink}
\usepackage{booktabs}
\usepackage{microtype}
\usepackage{xcolor}
\graphicspath{{figures/}}
\allowdisplaybreaks

\newcommand{\dd}{\mathrm{d}}
\newcommand{\Qcal}{\mathcal{Q}}
\newcommand{\Qbar}{\overline{Q}}
\newcommand{\alphaloc}{\alpha_{\mathrm{loc}}}
\newcommand{\alphatot}{\alpha_{\phi}}
\newcommand{\alphabg}{\alpha_{\phi}^{\mathrm{bg}}}
\newcommand{\Order}{\mathcal{O}}
\newcommand{\bKR}{b_{\mathrm{KR}}}

\begin{document}

\title{Chromatic Weak Lensing by Charged Black Holes with Two Lorentz-Violating Kalb-Ramond Couplings}

\author{Ali \"Ovg\"un \orcidlink{0000-0002-9889-342X}}
\email{ali.ovgun@emu.edu.tr}
\affiliation{Physics Department, Eastern Mediterranean University, Famagusta, Cyprus}

\author{Reggie C. Pantig \orcidlink{0000-0002-3101-8591}} 
\email{rcpantig@mapua.edu.ph}
\affiliation{Physics Department, School of Foundational Studies and Education, Map\'ua University, 658 Muralla St., Intramuros, Manila 1002  Philippines.}

\author{Grigoris Panotopoulos \orcidlink{0000-0003-1449-9108}} \email{grigorios.panotopoulos@ufrontera.cl}
\affiliation{Departamento de Ciencias F{\'i}sicas, Universidad de La Frontera, Casilla 54-D, 4811186 Temuco, Chile.}

\begin{abstract}
We study weak gravitational lensing and steady spherical test-fluid accretion by a static charged black hole in a Lorentz-violating Kalb–Ramond background with two nonminimal curvature couplings, assuming minimally coupled probe radiation. Using the Gauss–Bonnet theorem with the correct boundary term and perturbed ray boundary, we obtain the complete local deflection angle through second post-Minkowskian order.  In a homogeneous cold plasma, the mass and charge sectors acquire different frequency dependences, producing distinct chromatic signatures. The two Lorentz-violating couplings are also separated: one controls the conical geometry and mass normalization, while the other first enters through the effective charge. For neutral adiabatic accretion, we derive the conserved fluxes, Bernoulli relation, Hamiltonian flow, and sonic-point conditions. These results disentangle local, global, dispersive, and accretion effects and establish the calibrations required for phenomenological constraints.

\end{abstract}


\maketitle

\section{Introduction}
\label{sec:introduction}

Possible violations of local Lorentz invariance arise in several approaches to quantum gravity and high-energy physics. Spontaneous Lorentz-symmetry breaking was first shown to emerge naturally in string field theory when tensor fields acquire nonvanishing vacuum expectation values (VEVs) \cite{KosteleckySamuel1989,Kostelecky:1989jw}. Planck-scale modifications of particle propagation have also been investigated in loop quantum gravity \cite{Alfaro2002}, while noncommutative field theories generate effective Lorentz-violating operators at low energies \cite{Carroll:2001ws}. These developments are systematically organized within the gravitational sector of the Standard-Model Extension, which provides a covariant effective-field-theory description of Lorentz violation in gravity and matter \cite{Kostelecky:2003fs}.

Tensor fields with nonzero VEVs offer a particularly economical realization of spontaneous Lorentz-symmetry breaking. In addition to vector ``bumblebee'' models, antisymmetric rank-two fields admit nontrivial vacuum structures, propagating modes, and curvature couplings that cannot generally be reduced to those of a vector condensate. The spontaneous breaking mechanism for antisymmetric tensors and its field-theoretic consequences were developed in~\cite{Higashijima:2001sq,AltschulBaileyKostelecky2010}, while the corresponding propagator structure and physical degrees of freedom were analyzed in~\cite{Maluf:2018jwc}. Antisymmetric tensors can also couple directly to electromagnetic fields, potentially producing parity-sensitive or polarization-dependent propagation effects \cite{Majumdar:1999jd}. These properties make tensor-induced Lorentz violation relevant both to the construction of modified compact objects and to the propagation of observable fields in their vicinity.

The related bumblebee framework has developed into an extensive laboratory for black-hole and strong-gravity phenomenology. Exact static and rotating solutions have been obtained for timelike, spacelike, and lightlike vector VEVs, including electrically charged geometries whose asymptotic structure can differ from that of the standard Reissner--Nordstr\"om spacetime \cite{Liu2026BumblebeeExact,Liu2025ChargedBumblebee}. Their thermodynamic interpretation has also revealed that the physical mass, entropy, and first-law variables can depend nontrivially on the normalization of the Lorentz-violating background \cite{An2024Bumblebee}. These results emphasize that observable predictions should be expressed in terms of consistently normalized conserved quantities rather than metric integration constants alone.

Perturbations provide complementary information about the dynamical viability of such geometries. Quasinormal spectra have been calculated for slowly rotating Einstein-bumblebee black holes \cite{Liu2023BumblebeeQNM}, massive scalar perturbations \cite{Deng2025Bumblebee}, charged spherical configurations \cite{Li2025BumblebeeQNM}, and black holes containing a cosmological constant and a global monopole \cite{Singh2025}. Lorentz violation can break the isospectrality between different perturbative sectors \cite{Liu2024Isospectrality}, although an appropriate treatment of the coupled gravitational-bumblebee system can also yield exact decoupling relations and corresponding isospectral structures \cite{Liu2026Decoupling}. Beyond isolated black holes, the bumblebee model has been constrained through cosmological perturbations and gravitational-wave propagation \cite{Lai2026}, while its implications for quantum-correlation harvesting and acceleration radiation have been investigated in~\cite{Liu2025BTZ,Tang2025}. Together, these studies demonstrate that the physical content of Lorentz-violating backgrounds depends not only on the metric but also on the coupled perturbative and matter sectors.

For an antisymmetric Kalb-Ramond background, the first static black-hole solutions established that a radial tensor VEV can generate Schwarzschild-like geometries with modified asymptotics \cite{Lessa:2019bgi,Yang:2023wtu}. Subsequent work produced neutral \cite{Liu:2024gxr}, electrically charged \cite{Duan:2023qsg}, and more general exact black-hole branches \cite{Liu:2025KR}. The inclusion of both electric and magnetic charges has led to dyonic configurations \cite{Lin:2026KR}, while the most recent charged construction retains the combined effects of the two nonminimal curvature operators adopted in the present model \cite{Yang:2026KR}. Kalb-Ramond condensates can also support nontrivial horizonless geometries, including traversable wormholes obtained in minimally and curvature-coupled settings \cite{Lessa:2020imi,Maluf:2021eyu}. This variety of solutions illustrates how the tensorial character of the Lorentz-breaking vacuum can influence causal structure, asymptotic geometry, and conserved charges.

The observational and perturbative properties of Kalb-Ramond compact objects have consequently received increasing attention \cite{Shodikulov:2025xax,Nengroo:2026iju,Battista:2026rtl,Murodov:2026vmd,Battista:2026nsx,Kumar:2020hgm,Vagnozzi:2022moj}. Particle motion and weak gravitational lensing were studied in~\cite{Atamurotov:2022slw}, while black-hole shadows and quasinormal frequencies were used to characterize antisymmetric-tensor corrections in~\cite{AraujoFilho:2023qea}. Shadow formation around slowly rotating configurations was investigated in~\cite{Liu:2024yhu}. Quasinormal modes of electrically charged Kalb-Ramond black holes have also been considered \cite{GuChargedKRQNM}, and more complete analyses now include scalar, electromagnetic, and gravitational perturbations of slowly rotating solutions \cite{Deng:2025KR}. Related tensor-induced effects have additionally been explored for rotating BTZ backgrounds \cite{Xia:2025KR}. These results show that geodesic observables, wave dynamics, and electromagnetic propagation can probe different combinations of the Lorentz-violating parameters.

Against this background, the two-coupling charged solution of~\cite{Yang:2026KR} provides a timely setting in which to separate local curvature effects from global asymptotic geometry. Unlike a simple one-parameter deformation of Reissner--Nordstr\"om spacetime, the adopted branch contains distinct coupling combinations in its asymptotic normalization and charge sector. A consistent lensing analysis must therefore identify the locally normalized mass and charge, retain the non-Euclidean angular identification of the far region, and specify which electromagnetic polarization or probe field is being followed. The purpose of the present work is to perform this separation explicitly and to determine how the two Lorentz-violating directions enter vacuum and dispersive weak-deflection observables.

Weak lensing by a compact object is controlled not only by the curvature accumulated near the ray but also by the geometry used to compare the incoming and outgoing directions.  The familiar statement that a deflection angle vanishes as the impact parameter tends to infinity presupposes an asymptotically Euclidean optical surface.  When the far region instead approaches a cone, local curvature bending still decays with distance, but a global angular mismatch remains.  Separating these two effects is essential before a Lorentz-violating parameter can be associated with an observable image displacement.

Antisymmetric Kalb-Ramond two-forms provide a tensorial realization of spontaneous Lorentz-symmetry breaking.  They arise in the massless bosonic sector of string-inspired theories, and a nonzero vacuum expectation value of $B_{\mu\nu}$ selects preferred spacetime directions \cite{KalbRamond1974,KosteleckySamuel1989,AltschulBaileyKostelecky2010}.  Nonminimal contractions of this vacuum tensor with curvature then modify the gravitational field equations.  Static neutral and charged black-hole solutions have been constructed in this setting \cite{Yang:2023wtu,Duan:2023qsg,Liu:2025KR}.  The solution studied here retains both independent curvature couplings and therefore contains more information than the one-parameter truncations commonly used in optical analyses \cite{Yang:2026KR}.

The Gauss-Bonnet method is well suited to this problem because it exposes both bulk optical curvature and boundary geometry.  For an asymptotically Euclidean optical manifold, the Gibbons-Werner construction expresses the bending as a curvature integral over the region exterior to the ray \cite{GibbonsWerner2008,Werner2012}.  Finite-distance or non-asymptotically flat spacetimes require a careful treatment of the boundary terms and of the angle definition \cite{IshiharaEtAl2016}.  A cold nonmagnetized plasma can be incorporated through a frequency-dependent optical metric, allowing the same geometric strategy to describe dispersive propagation \cite{CrisnejoGallo2018,PerlickTsupkoBisnovatyiKogan2015,BisnovatyiKoganTsupko2010}.

Our calculation follows the complementary Gauss-Bonnet, geodesic, and plasma program used for charged nonlinear-electrodynamic black holes in~\cite{FuZhaoLiu2021}, but the present spacetime requires two substantive changes.  First, the metric does not approach Minkowski spacetime in its original coordinates, so the large-circle contribution cannot be imported from an asymptotically flat calculation.  Second, a straight zeroth-order ray is insufficient at second post-Minkowskian (2PM) order: it reproduces the leading mass and charge contributions but misses the part of the $M^{2}/b^{2}$ coefficient generated by the displaced integration boundary.

Optical properties of one-parameter Kalb-Ramond black holes have already been investigated through shadows, null geodesics, quasinormal modes, and weak deflection \cite{JuniorEtAl2024,AraujoFilhoEtAl2024,TangLan2025,PantigOvgunRincon2025}.  The unresolved question for the two-coupling solution is not merely whether the angle changes, but \emph{which physical sector carries each coupling}.  We answer that question by distinguishing four contributions: a constant conical term, a $b^{-1}$ mass term, a $b^{-2}$ charge term, and frequency-dependent plasma weights.  This hierarchy also makes clear where parameter degeneracies arise and which signatures could, in principle, separate them.

The main results are as follows.  After canonical normalization, the local null orbit is exactly of Reissner--Nordstr\"om form, whereas the global azimuth retains a nonstandard period.  The complete vacuum 2PM angle obtained from the Gauss-Bonnet theorem agrees with an independent null-geodesic expansion, and the exact scattering integral provides a numerical benchmark beyond the weak-field truncation.  A homogeneous plasma assigns different frequency weights to the leading mass, second-order mass, and charge sectors. To first order about the general-relativistic limit and at fixed metric integration constants, $l_2$ changes the cone and normalized mass, while $l_1$ changes the effective charge.  As a complementary matter diagnostic, steady spherical accretion yields the correct baryon and
Killing-energy integrals, the associated Hamiltonian and sonic conditions, and a regular cold-dust benchmark.  In normalized variables its local sonic structure is Reissner--Nordstr\"om-like, whereas the conical factor enters the integrated accretion rate through the area of the symmetry spheres.  We support the optical and accretion results with exact numerical checks and state explicitly the normalization and test-probe assumptions under which they apply.

 Section~\ref{sec:model} summarizes the field-theory branch and its approximations.  Section~\ref{sec:cone} constructs the normalized optical cone and separates local, background-subtracted, and topology-retaining coordinate angles.  The vacuum Gauss-Bonnet and null-geodesic calculations are given in Secs.~\ref{sec:gbt} and \ref{sec:geodesic}, and Sec.~\ref{sec:plasma} develops the homogeneous-plasma extension.  Numerical validation and parameter response are presented in Sec.~\ref{sec:numerics}. Section~\ref{sec:accretion} derives the steady spherical fluid equations and the cold-dust benchmark.  The physical interpretation, limitations, and main conclusions are given in Secs.~\ref{sec:discussion} and \ref{sec:conclusions}.  We use $G=c_{\rm light}=1$ and signature $(-,+,+,+)$.

\section{Field-theory branch and charged geometry}
\label{sec:model}

\subsection{Action, vacuum configuration, and approximations}

The gravitational model contains a self-interacting two-form together with the two independent curvature contractions allowed by the chosen Kalb-Ramond background.  In the normalization of the source solution, the action is \cite{AltschulBaileyKostelecky2010}
\begin{equation}
\begin{aligned}
S={}&\int \dd^{4}x\sqrt{-g}\bigg[
 \frac{R-2\Lambda}{2\kappa}
 -\frac{1}{12}H^{\mu\nu\rho}H_{\mu\nu\rho}-V(X) \\
&+\frac{\xi_{1}}{2\kappa}B^{\mu\nu}B_{\mu\nu}R
 +\frac{\xi_{2}}{2\kappa}B^{\rho\mu}B^{\nu}{}_{\mu}R_{\rho\nu} \\
&+\frac{1}{2\kappa}\mathcal{L}_{M}\bigg].
\end{aligned}
\label{eq:action}
\end{equation}
Here $\kappa=8\pi$, $X$ measures displacement from the symmetry-breaking vacuum, and $\xi_{1}$ and $\xi_{2}$ multiply genuinely distinct curvature operators.  The first operator changes the scalar-curvature sector, whereas the second projects the Ricci tensor along directions selected by the antisymmetric vacuum tensor; retaining both is therefore physically more informative than absorbing all Lorentz violation into a single metric parameter.

The gauge-invariant three-form and the electromagnetic matter sector are defined by \cite{Duan:2023qsg,Higashijima:2001sq}
\begin{align}
H_{\mu\nu\rho}&=\partial_{\mu}B_{\nu\rho}
 +\partial_{\nu}B_{\rho\mu}
 +\partial_{\rho}B_{\mu\nu},
\label{eq:Hdef}\\
\mathcal{L}_{M}&=-\frac{1}{2}F^{\mu\nu}F_{\mu\nu}
 -\eta B^{\alpha\beta}B^{\gamma\rho}
 F_{\alpha\beta}F_{\gamma\rho}.
\label{eq:matterL}
\end{align}
The algebraic $B^{2}F^{2}$ term is required by the charged branch under consideration.  The unusual normalization of the Maxwell term is kept fixed, so the charge parameter below must be interpreted in the same convention as Eq.~\eqref{eq:matterL}.

We work on the radial pseudo-electric vacuum configuration, for which $B_{tr}=-B_{rt}$ is the only independent component and \cite{Maluf:2018jwc,Lessa:2019bgi}
\begin{equation}
B^{\mu\nu}B_{\mu\nu}=-\bKR^{2},
\qquad
H_{\mu\nu\rho}=0,
\qquad
V=V'=0.
\label{eq:vacuumbranch}
\end{equation}
These conditions place the two-form at the minimum of its potential and make the background static.  Our lensing analysis treats this vacuum tensor as a fixed external background: perturbations of $B_{\mu\nu}$, plasma backreaction on the metric, and mode conversion between electromagnetic and two-form fluctuations are outside the present approximation.

It is convenient to introduce the dimensionless curvature couplings
\begin{equation}
 l_{1}=\bKR^{2}\xi_{1},
 \qquad
 l_{2}=\bKR^{2}\xi_{2}.
\label{eq:l1l2}
\end{equation}
On the exact electrically charged branch, the algebraic electromagnetic coupling is not independent but satisfies
\begin{equation}
 \eta=\frac{l_{2}}{4\bKR^{2}(1-l_{1})}.
\label{eq:etarelation}
\end{equation}
Thus $l_{1}$ and $l_{2}$ remain independent curvature parameters, while $\eta$ is fixed by the field equations once the branch is selected.

\subsection{Asymptotically conical black-hole branch}

For the quadratic potential on its vacuum branch and $\Lambda=0$, the static, spherically symmetric line element can be written as \cite{Yang:2026KR}
\begin{equation} 
 \dd s^{2}=-F(r)\dd t^{2}+\frac{\dd r^{2}}{F(r)}
 +r^{2}\left(\dd\theta^{2}+\sin^{2}\theta\,\dd\phi^{2}\right).
\label{eq:originalmetric}
\end{equation}
The metric function separates naturally into an asymptotic constant, a mass term, and a charge term, 
\begin{align}
 F(r)&=c-\frac{2M}{r}+\frac{\Qcal^{2}}{r^{2}},
\label{eq:F}\\
 c&=\frac{1+l_{1}}{1+l_{1}-l_{2}/2},
\qquad
 \Qcal^{2}=\frac{(1-l_{1})Q^{2}}
 {(1-l_{1}-l_{2}/2)^{2}}.
\label{eq:cQ}
\end{align}
The parameter $c$ controls the far-field normalization and global angular structure, while $\Qcal$ is the charge combination that appears in the metric.  The charge term in the metric falls off as $r^{-2}$ and therefore does not alter the leading conical asymptotics.

The branch describes a positive-mass black hole rather than a naked singularity when the asymptotic time direction is timelike, the effective charge squared is nonnegative, and the horizon discriminant is nonnegative.  These requirements are
\begin{equation}
 M>0,
 \qquad
 c>0,
 \qquad
 \Qcal^{2}\geq0,
 \qquad
 M^{2}\geq c\Qcal^{2}.
\label{eq:domain}
\end{equation}
Within this domain, the inner and outer horizons are
\begin{equation}
 r_{\pm}=\frac{M\pm\sqrt{M^{2}-c\Qcal^{2}}}{c}.
\label{eq:horizons}
\end{equation}
The weak-lensing trajectories studied below lie outside $r_{+}$ and, more restrictively, have impact parameters above the critical value associated with the unstable photon sphere.

\section{Canonical normalization, optical cone, and observable hierarchy}
\label{sec:cone}

\subsection{Locally normalized geometry}

Because $F(r)\rightarrow c$ rather than unity, the coordinate $t$ is not the proper time of a static observer in the asymptotic region.  A simultaneous rescaling of time and radius removes this local normalization ambiguity:
\begin{equation}
 T=\sqrt{c}\,t,
 \qquad
 R=\frac{r}{\sqrt{c}}.
\label{eq:TR}
\end{equation}
Substitution into Eq.~\eqref{eq:originalmetric} gives
\begin{equation}
 \dd s^{2}=-A(R)\dd T^{2}+\frac{\dd R^{2}}{A(R)}
 +cR^{2}\dd\Omega^{2},
\label{eq:normalizedmetric}
\end{equation}
where the locally measured potential is
\begin{equation}
 A(R)=1-\frac{2\mu}{R}+\frac{\Qbar^{2}}{R^{2}},
 \qquad
 \mu=\frac{M}{c^{3/2}},
 \qquad
 \Qbar^{2}=\frac{\Qcal^{2}}{c^{2}}.
\label{eq:effectiveparameters}
\end{equation}
The local dynamics therefore depends on the normalized mass $\mu$ and charge $\Qbar$, not directly on the integration constants $M$ and $Q$.  This distinction is important in any phenomenological comparison because the spacetime has no standard asymptotically flat ADM normalization in the original coordinates.

On the equatorial plane, the azimuthal coordinate
\begin{equation}
 \varphi=\sqrt{c}\,\phi
\label{eq:varphi}
\end{equation}
puts the metric into the local Reissner--Nordstr\"om form,
\begin{equation}
 \dd s^{2}_{\theta=\pi/2}=-A\dd T^{2}+A^{-1}\dd R^{2}
 +R^{2}\dd\varphi^{2}.
\label{eq:localRN}
\end{equation}
This equivalence is local rather than global: because $\phi$ has period $2\pi$, the normalized angle $\varphi$ has period $2\pi\sqrt{c}$.  Hence the far-field equatorial surface is a cone, even though Eq.~\eqref{eq:localRN} looks locally flat as $R\rightarrow\infty$.

The conserved energies and angular momenta in the two coordinate systems are related in the same way.  If $E=F\dot t$, $L=r^{2}\dot\phi$, $\mathcal{E}=A\dot T$, and $J=R^{2}\dot\varphi$, then
\begin{equation}
 E=\sqrt{c}\,\mathcal{E},
 \qquad
 L=\sqrt{c}\,J,
 \qquad
 b\equiv\frac{L}{E}=\frac{J}{\mathcal{E}}.
\label{eq:bInvariant}
\end{equation}
The impact parameter $b$ is therefore invariant under the normalization.  It is the geometric impact parameter on the locally Euclidean covering plane: the zeroth-order closest approach is $R_{\min}^{(0)}=b$, whereas the original areal coordinate gives $r_{\min}^{(0)}=\sqrt{c}\,b$.  This allows the analytic and numerical calculations to be performed in the locally canonical variables without changing the physical labeling of a vacuum scattering orbit. The equality $b=L/E=J/\mathcal{E}$ is the vacuum-photon relation.  When the
ray has asymptotic speed $v<1$, as for a massive particle or a photon in a
homogeneous plasma, we define the same geometric impact parameter by
$b=J/(\mathcal{E}v)$, so that $J/\mathcal{E}=vb$.

\subsection{Local, background-subtracted, and coordinate angular changes}

At finite source and receiver radii, it is useful to define the local angle
before taking the asymptotic limit.  Let $\Psi_{S}$ and $\Psi_{R}$ denote the
trajectory angles measured by static observers, and let
$\varphi_{RS}=\varphi_{R}-\varphi_{S}$.  We define
\begin{equation}
 \alpha_{\mathrm{loc}}(R_{S},R_{R})
 =\Psi_{R}-\Psi_{S}+\varphi_{RS}.
\label{eq:finiteLocalAngle}
\end{equation}
The corresponding angular change written with the original azimuth is
\begin{equation}
 \alpha_{\phi}(r_{S},r_{R})
 =\Psi_{R}-\Psi_{S}+\phi_{RS}
 =\alpha_{\mathrm{loc}}
 +\left(c^{-1/2}-1\right)\varphi_{RS},
\label{eq:finiteAngleConversion}
\end{equation}
where $\varphi_{RS}=\sqrt{c}\,\phi_{RS}$.  This form keeps the local bending
separate from the global angular identification before any weak-field or
far-distance limit is imposed.  It follows the finite-distance angle logic
used for Schwarzschild-like geometries in Ref.~\cite{Pantig:2024kqy}, with
the asymptotic normalization performed first for the present charged cone.
In the limit $R_{S},R_{R}\rightarrow\infty$, Eq.~\eqref{eq:finiteAngleConversion}
reduces to Eqs.~\eqref{eq:angledefs} and \eqref{eq:anglerelation}.

A conical spacetime admits several useful, but physically distinct, angular
comparisons.  For total coordinate changes $\Delta\phi$ and
$\Delta\varphi=\sqrt{c}\,\Delta\phi$, we define
\begin{equation}
 \alphaloc=\Delta\varphi-\pi,
 \qquad
 \alphabg=\Delta\phi-\frac{\pi}{\sqrt{c}},
 \qquad
 \alphatot=\Delta\phi-\pi.
\label{eq:angledefs}
\end{equation}
The local angle $\alphaloc$ compares the ray with a straight trajectory on
the unwrapped asymptotic cone, while $\alphabg$ is the same
background-subtracted bending expressed in the original coordinate.  The
quantity $\alphatot$ is instead the coordinate angular excess relative to
the Euclidean value $\pi$; it retains the global mismatch produced by
identifying the cone with the original $2\pi$ coordinate range and is not a
locally measured deflection angle.

The three conventions obey the exact relations
\begin{equation}
 \alphabg=\frac{\alphaloc}{\sqrt{c}},
 \qquad
 \alphatot=\pi\left(c^{-1/2}-1\right)+\alphabg.
\label{eq:anglerelation}
\end{equation}
The first term is independent of $b$ and survives even when $M=Q=0$.  It is therefore not a residual gravitational force.  It represents the global angular structure and must be incorporated through a lens equation or source-observer construction adapted to the cone; by contrast, $\alphaloc\rightarrow0$ as $b\rightarrow\infty$.

\begin{figure}[t]
 \includegraphics[width=\columnwidth]{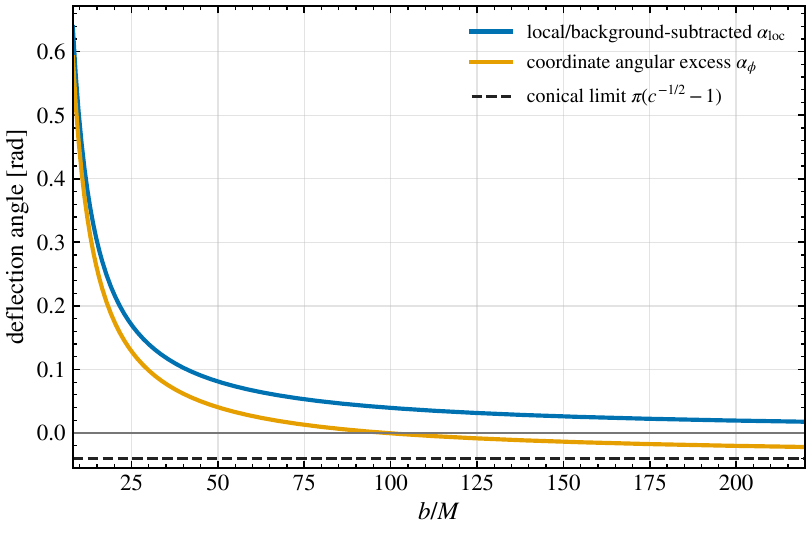}
 \caption{Local/background-subtracted angle and coordinate angular excess at 2PM order for $l_{1}=0$, $l_{2}=0.05$, $M=1$, and $Q/M=0.6$.  The horizontal dashed line is the pure conical term in Eq.~\eqref{eq:anglerelation}.  The parameter values are chosen to make the separation visible and are not used as phenomenological bounds.}
 \label{fig:angleConvention}
\end{figure}

Figure~\ref{fig:angleConvention} displays this distinction before any weak-field expansion is interpreted observationally.  The local angle decays, whereas the coordinate angular excess approaches the conical constant.

\subsection{Linear coupling map and thin-lens observables}

The physical separation between the two Lorentz-violating couplings is already visible in the small-coupling expansion.  To first order in $l_{1}$ and $l_{2}$,
\begin{align}
 c&=1+\frac{l_{2}}{2}+\Order(l_{i}^{2}),
\label{eq:cLinear}\\
 \mu&=M\left(1-\frac{3l_{2}}{4}\right)+\Order(l_{i}^{2}),
\label{eq:muLinear}\\
 \Qbar^{2}&=Q^{2}(1+l_{1})+\Order(l_{i}^{2}).
\label{eq:qLinear}
\end{align}
The cancellation of $l_{1}$ from $c$ and $\mu$ at linear order means that $l_{2}$ controls the leading achromatic mass and cone sectors.  The coupling $l_{1}$ first appears through the charge term, which is both subleading in $1/b$ and absent when $Q=0$.

For the background-subtracted local angle, a standard small-angle lens equation may be used as a diagnostic,
\begin{equation}
 \beta_{s}=\theta-\frac{D_{LS}}{D_{S}}
 \alphaloc(D_{L}|\theta|)\,\mathrm{sgn}(\theta),
\label{eq:lensequationlocal}
\end{equation}
where $\beta_{s}$ and $\theta$ are the unlensed source and image angles and $D_{L}$, $D_{S}$, and $D_{LS}$ are angular-diameter distances.  Equation~\eqref{eq:lensequationlocal} is deliberately written for $\alphaloc$; inserting the topology-retaining coordinate excess $\alphatot$ into a flat-background lens equation would misinterpret the conical offset.

Keeping only the leading vacuum mass term gives the local Einstein-ring scale
\begin{equation}
 \theta_{E}^{\rm vac}=\left(\frac{4\mu D_{LS}}{D_{L}D_{S}}\right)^{1/2}.
\label{eq:EinsteinVac}
\end{equation}
At fixed $M$, Eq.~\eqref{eq:muLinear} gives $\Delta\theta_{E}/\theta_{E}=-3l_{2}/8+\Order(l_{i}^{2})$.  This is not by itself a clean constraint on $l_{2}$ because the same observable also calibrates the mass; the useful information comes from combining different impact-parameter or frequency scalings.

Table~\ref{tab:hierarchy} summarizes the hierarchy that guides the rest of the paper.  It distinguishes a global $b^{0}$ contribution from local $b^{-1}$ and $b^{-2}$ curvature terms and shows where chromatic information enters.
\begin{table*}[t]
\caption{Physical hierarchy of the weak-lensing contributions.  ``Coupling direction'' refers to the first nonvanishing term in a small-$(l_{1},l_{2})$ expansion.}
\label{tab:hierarchy}
\small
\begin{tabular}{@{}p{0.13\textwidth}p{0.09\textwidth}p{0.16\textwidth}p{0.23\textwidth}p{0.28\textwidth}@{}}
\toprule
Contribution & Scaling & Coupling direction & Vacuum behavior & Homogeneous-plasma behavior \\
\midrule
Global cone & $b^{0}$ & $l_{2}$ & Achromatic angular identification. & Unchanged by the constant asymptotic refractive factor. \\
Local mass & $b^{-1}$ & $l_{2}$ through $\mu$ & Attractive and dominant. & Enhanced by $\tfrac12(1+v^{-2})$. \\
2PM mass & $b^{-2}$ & $l_{2}$ through $\mu^{2}$ & Attractive. & Enhanced by $\tfrac15(1+4v^{-2})$. \\
2PM charge & $b^{-2}$ & $l_{1}$ through $\Qbar^{2}$ & Reduces the RN bending. & Enhanced in magnitude by $\tfrac13(1+2v^{-2})$. \\
\bottomrule
\end{tabular}
\end{table*}

\section{Vacuum deflection from the Gauss-Bonnet theorem}
\label{sec:gbt}

\subsection{Optical curvature and its physical content}

For equatorial null rays in the normalized geometry, setting $\dd s^{2}=0$ solves the coordinate time in terms of a two-dimensional optical line element,
\begin{equation}
 \dd T^{2}=\dd\sigma^{2}
 =\frac{\dd R^{2}}{A^{2}(R)}+\frac{R^{2}}{A(R)}\dd\varphi^{2}.
\label{eq:opticalmetric}
\end{equation}
Spatial projections of null geodesics are geodesics of this optical metric.  The lensing problem can therefore be expressed as a statement about the intrinsic curvature of the optical surface plus the geometry of its boundary.

Introducing the optical radial coordinate $\dd R_{\star}=\dd R/A$ and circumference function $f_{\rm opt}=R/\sqrt{A}$, the Gaussian curvature is
\begin{equation}
 K=-\frac{1}{f_{\rm opt}}\frac{\dd^{2}f_{\rm opt}}{\dd R_{\star}^{2}}
 =\frac{AA''}{2}-\frac{(A')^{2}}{4}.
\label{eq:Kcompact}
\end{equation}
This compact expression makes the calculation insensitive to a particular embedding of the optical surface.  For the effective potential in Eq.~\eqref{eq:effectiveparameters}, it evaluates exactly to
\begin{equation}
 K=-\frac{2\mu}{R^{3}}
 +\frac{3(\mu^{2}+\Qbar^{2})}{R^{4}}
 -\frac{6\mu\Qbar^{2}}{R^{5}}
 +\frac{2\Qbar^{4}}{R^{6}}.
\label{eq:Kexact}
\end{equation}
The leading mass term is negative and produces attractive bending through $-\iint K\dd S$.  The first charge-dependent term is positive and proportional to $R^{-4}$.  It therefore reduces the Reissner--Nordstr\"om deflection at fixed mass.

The corresponding surface element is
\begin{equation}
 \dd S=\frac{R}{A^{3/2}}\dd R\dd\varphi.
\label{eq:dS}
\end{equation}
Both $K$ and $\dd S$ are required at consistent perturbative order: expanding only the curvature while keeping an inconsistent area element would alter the 2PM coefficient.

\subsection{Conical boundary term and direct coordinate formula}

Let $D_{\infty}$ be the simply connected region outside the light ray,
closed by a circular arc in the asymptotic domain.  It is useful first to
apply the Gauss--Bonnet theorem directly in the original coordinates.  On
the equatorial plane the corresponding optical metric is
\begin{equation}
 \dd t^{2}=\dd\sigma_{(t)}^{2}
 =\frac{\dd r^{2}}{F^{2}(r)}+\frac{r^{2}}{F(r)}\dd\phi^{2}.
\label{eq:originalOpticalMetric}
\end{equation}
For a circle $C_{r}:r={\rm const}$, its geodesic-curvature measure is
\begin{equation}
 \left.\kappa_{g}\dd\ell\right|_{C_{r}}
 =\left(\sqrt{F}-\frac{rF'}{2\sqrt{F}}\right)\dd\phi
 \xrightarrow[r\rightarrow\infty]{}\sqrt{c}\,\dd\phi
 =\dd\varphi.
\label{eq:circleboundary}
\end{equation}
This is the essential conical correction.  Replacing the limiting measure
by $\dd\phi$ would silently impose $c=1$ and mix local curvature bending
with the global angular identification.

The light ray is an optical geodesic and hence makes no geodesic-curvature
contribution.  The two asymptotic jump angles sum to $\pi$, so the
Gauss--Bonnet theorem gives
\begin{equation}
 I_{\gamma}+\sqrt{c}\,\Delta\phi=\pi,
 \qquad
 I_{\gamma}\equiv\iint_{D_{\infty}}K_{(t)}\,\dd S_{(t)}.
\label{eq:directGBT}
\end{equation}
Consequently, the coordinate angular excess defined in
Eq.~\eqref{eq:angledefs} obeys the direct conical formula
\begin{equation}
 \alphatot=\pi\left(c^{-1/2}-1\right)
 -\frac{I_{\gamma}}{\sqrt{c}}.
\label{eq:directConicalAngle}
\end{equation}
This is the analogue of the cosmic-string Gauss--Bonnet formula, with
$\sqrt{c}$ playing the role of the asymptotic cone parameter.

The curvature and area element of Eq.~\eqref{eq:originalOpticalMetric} are
\begin{equation}
 K_{(t)}=\frac{FF''}{2}-\frac{(F')^{2}}{4},
 \qquad
 \dd S_{(t)}=\frac{r}{F^{3/2}}\dd r\dd\phi.
\label{eq:originalKdS}
\end{equation}
The constant rescaling $T=\sqrt{c}\,t$ changes $K$ and $\dd S$
separately but leaves their product invariant; the subsequent coordinate
change $(r,\phi)\mapsto(R,\varphi)$ therefore gives
$K_{(t)}\dd S_{(t)}=K\dd S$ on the same geometric domain.  It follows that
\begin{equation}
 \alphaloc=-I_{\gamma}=-\iint_{D_{\infty}}K\,\dd S,
\label{eq:GBTalpha}
\end{equation}
and Eq.~\eqref{eq:directConicalAngle} is exactly equivalent to the angle
conversion in Eq.~\eqref{eq:anglerelation}.

There is one further geometrical point that is important in perturbation
theory.  The undeflected ray is straight in the unwrapped angle
$\varphi$, not in $\phi$.  Its original-coordinate form is therefore
\begin{equation}
 r_{0}(\phi)=\frac{\sqrt{c}\,b}{\sin(\sqrt{c}\,\phi)},
 \qquad 0<\phi<\frac{\pi}{\sqrt{c}},
\label{eq:conicalStraightRay}
\end{equation}
rather than $r=b/\sin\phi$.  This distinction is immaterial when $c=1$
but is essential when the conical contribution is retained exactly.

Equations~\eqref{eq:directGBT}--\eqref{eq:conicalStraightRay} rely on the
fact that the present optical geometry tends to a cone with a constant
opening parameter.  They are not a universal prescription for every
non-asymptotically-flat spacetime.  When no such conical limit exists, the
source and receiver should be kept at finite distance and the invariant
angle $\Psi_{R}-\Psi_{S}+\varphi_{RS}$ should be compared with the chosen
reference background before any asymptotic limit is considered
\cite{IshiharaEtAl2016}.

\subsection{Complete second-post-Minkowskian result}

We count $\mu/b=\Order(\epsilon)$ and $\Qbar^{2}/b^{2}=\Order(\epsilon^{2})$.  Expanding the curvature density through second order gives
\begin{equation}
 K\dd S=\left[-\frac{2\mu}{R^{2}}
 +\frac{3(\Qbar^{2}-\mu^{2})}{R^{3}}\right]
 \dd R\dd\varphi+\Order(\epsilon^{3}).
\label{eq:KdSvacexp}
\end{equation}
The $\mu^{2}$ term in this density is only one source of the 2PM mass correction.  A second contribution arises because the lower boundary of the curvature integral is itself displaced by the first-order gravitational field.

The required first-order orbit is
\begin{equation}
 U_{\gamma}(\varphi)=\frac{\sin\varphi}{b}
 +\frac{\mu}{b^{2}}\left(1+\cos^{2}\varphi\right)
 +\Order(\epsilon^{2}),
 \qquad U\equiv R^{-1}.
\label{eq:orbitfirstvac}
\end{equation}
The first term is the straight ray in polar coordinates, while the second shifts the integration boundary toward the lens.  Omitting this shift yields an incomplete $\mu^{2}/b^{2}$ coefficient even though the curvature density itself has been expanded to second order.  Equation~\eqref{eq:orbitfirstvac} uses a closest-approach-symmetric representative, so its asymptotic roots are displaced from $0$ and $\pi$.  An incoming-asymptote-anchored representative differs only by a homogeneous term proportional to $\cos\varphi$, whose integral over $[0,\pi]$ vanishes; corrections from the shifted endpoint intervals begin at 3PM order.

For comparison with the direct conical formula, the same two ingredients
in the original coordinates are
\begin{align}
 K_{(t)}\dd S_{(t)}={}&\left[-\frac{2M}{\sqrt{c}\,r^{2}}
 +\frac{3\Qcal^{2}}{\sqrt{c}\,r^{3}}
 -\frac{3M^{2}}{c^{3/2}r^{3}}\right]
 \dd r\dd\phi+\Order(\epsilon^{3}),
\label{eq:originalKdSexp}\\
 \frac{1}{r_{\gamma}(\phi)}={}&
 \frac{\sin(\sqrt{c}\,\phi)}{\sqrt{c}\,b}
 +\frac{M}{c^{2}b^{2}}
 \left[1+\cos^{2}(\sqrt{c}\,\phi)\right]
 +\Order(\epsilon^{2}).
\label{eq:originalOrbitFirst}
\end{align}
Using the background range $0<\phi<\pi/\sqrt{c}$ in the curvature
integral reproduces the normalized calculation below.  In particular, the
second term of Eq.~\eqref{eq:originalOrbitFirst} supplies the same missing
2PM boundary-displacement contribution as Eq.~\eqref{eq:orbitfirstvac}.

Substituting Eqs.~\eqref{eq:KdSvacexp} and \eqref{eq:orbitfirstvac} into Eq.~\eqref{eq:GBTalpha} gives
\begin{align}
 \alphaloc
 &=\int_{0}^{\pi}\left[
 2\mu U_{\gamma}
 -\frac{3}{2}(\Qbar^{2}-\mu^{2})U_{0}^{2}
 \right]\dd\varphi+\Order(\epsilon^{3})
 \nonumber\\
 &=\frac{4\mu}{b}
 +\frac{15\pi\mu^{2}}{4b^{2}}
 -\frac{3\pi\Qbar^{2}}{4b^{2}}
 +\Order(\epsilon^{3}),
\label{eq:alphavaclocal}
\end{align}
where $U_{0}=\sin\varphi/b$.  The three displayed contributions have distinct physical roles: the leading mass term dominates at large $b$, the positive 2PM mass term strengthens the attraction, and the negative charge term weakens it.  Details of the second-order integrations are collected in Appendix~\ref{app:gbt}.  Setting $l_{1}=l_{2}=0$ recovers the standard Reissner--Nordstr\"om expansion \cite{EiroaRomeroTorres2002,KeetonPetters2005}.

In terms of the original solution parameters, the same local result reads
\begin{equation}
 \alphaloc=\frac{4M}{c^{3/2}b}
 +\frac{15\pi M^{2}}{4c^{3}b^{2}}
 -\frac{3\pi\Qcal^{2}}{4c^{2}b^{2}}
 +\Order(\epsilon^{3}).
\label{eq:alphavacoriginal}
\end{equation}
This form shows explicitly that simply replacing $M$ and $Q$ in an asymptotically flat formula is not enough: the asymptotic normalization appears with different powers in the mass and charge sectors.

Restoring the original angular coordinate gives the coordinate angular excess
\begin{equation}
 \begin{aligned}
 \alphatot={}&\pi(c^{-1/2}-1)+\frac{4M}{c^{2}b}
 +\frac{15\pi M^{2}}{4c^{7/2}b^{2}}\\
 &-\frac{3\pi\Qcal^{2}}{4c^{5/2}b^{2}}
 +\Order(\epsilon^{3}).
 \end{aligned}
\label{eq:alphavaccoordinate}
\end{equation}
The first term must not be fitted as if it were another local $1/b$ force.  It is a global offset whose observable effect depends on how source and observer directions are defined on the cone.

Combining Eq.~\eqref{eq:alphavaclocal} with Eqs.~\eqref{eq:cLinear}-\eqref{eq:qLinear} gives the explicit linear-coupling response,
\begin{equation}
 \begin{aligned}
 \alphaloc={}&\frac{4M}{b}\left(1-\frac{3l_{2}}{4}\right)
 +\frac{15\pi M^{2}}{4b^{2}}
 \left(1-\frac{3l_{2}}{2}\right)\\
 &-\frac{3\pi Q^{2}}{4b^{2}}(1+l_{1})
 +\Order(l_{i}^{2},\epsilon^{3}).
 \end{aligned}
\label{eq:linearalpha}
\end{equation}
This equation makes the parameter identifiability transparent.  The leading effect of $l_{2}$ is degenerate with a mass renormalization, while $l_{1}$ can be accessed only through a charge-sensitive 2PM contribution or through another observable that independently fixes $\Qbar$.

\section{Null-geodesic calculation}
\label{sec:geodesic}

The geodesic method provides an independent check that does not rely on the optical-domain construction.  The equatorial null Lagrangian of the normalized metric is
\begin{equation}
 2\mathcal{L}=-A\dot T^{2}+A^{-1}\dot R^{2}
 +R^{2}\dot\varphi^{2}=0.
\label{eq:nullLagrangian}
\end{equation}
The null condition removes the affine-parameter normalization, leaving the orbit determined by the invariant ratio $b=J/\mathcal{E}$.

Using the two conserved quantities and $U=1/R$, the radial equation becomes
\begin{equation}
 \left(\frac{\dd U}{\dd\varphi}\right)^{2}
 =\frac{1}{b^{2}}-U^{2}+2\mu U^{3}-\Qbar^{2}U^{4}.
\label{eq:vacorbit}
\end{equation}
The first positive zero of the right-hand side is the inverse distance of closest approach.  Denoting it by $U_{m}$, it satisfies
\begin{equation}
 \frac{1}{b^{2}}-U_{m}^{2}+2\mu U_{m}^{3}
 -\Qbar^{2}U_{m}^{4}=0.
\label{eq:turningvac}
\end{equation}
For a scattering trajectory, this root is reached once and the ray returns to the asymptotic region; below the critical impact parameter, the turning point disappears and the ray is captured.

The exact background-subtracted scattering angle is therefore
\begin{equation}
 \alphaloc^{\rm exact}=2\int_{0}^{U_{m}}
 \frac{\dd U}
 {\sqrt{b^{-2}-U^{2}+2\mu U^{3}-\Qbar^{2}U^{4}}}-\pi.
\label{eq:exactvacintegral}
\end{equation}
This integral is evaluated numerically without a weak-field expansion and is used as the reference against which all truncated formulas are tested.

Differentiating Eq.~\eqref{eq:vacorbit} gives the Binet equation
\begin{equation}
 U''+U=3\mu U^{2}-2\Qbar^{2}U^{3}.
\label{eq:Binetvac}
\end{equation}
Its first iterate reproduces Eq.~\eqref{eq:orbitfirstvac}, confirming that the boundary correction used in the Gauss-Bonnet integral is the same physical displacement obtained directly from the orbit dynamics.

Expanding the exact integral at fixed invariant impact parameter gives
\begin{equation}
 \alphaloc^{\rm exact}=\frac{4\mu}{b}
 +\frac{15\pi\mu^{2}-3\pi\Qbar^{2}}{4b^{2}}
 +\frac{128\mu^{3}/3-16\mu\Qbar^{2}}{b^{3}}
 +\Order(\epsilon^{4}).
\label{eq:vac3pm}
\end{equation}
Agreement through 2PM order is a nontrivial joint check of the conical angle convention, the optical surface element, and the perturbed ray boundary.  The third-order term is used only as a convergence benchmark; the Gauss-Bonnet derivation in Sec.~\ref{sec:gbt} is consistently truncated at second order.

The outer photon sphere and vacuum critical impact parameter are determined by the local effective potential,
\begin{equation}
 R_{\rm ph}=\frac{3\mu+\sqrt{9\mu^{2}-8\Qbar^{2}}}{2},
 \qquad
 b_{c}=\frac{R_{\rm ph}}{\sqrt{A(R_{\rm ph})}}.
\label{eq:bc}
\end{equation}

The same critical orbit gives the shadow without introducing a new strong-field
calculation.  Transforming the outer photon sphere back to the original areal
coordinate gives
\begin{equation}
 r_{\mathrm{ph}}
 =\sqrt{c}\,R_{\mathrm{ph}}
 =\frac{3M+\sqrt{9M^{2}-8c\Qcal^{2}}}{2c}.
\label{eq:rphOriginal}
\end{equation}
The critical impact parameter can be written equivalently as
\begin{equation}
 b_{c}^{2}
 =\frac{r_{\mathrm{ph}}^{2}}{F(r_{\mathrm{ph}})}
 =\frac{r_{\mathrm{ph}}^{4}}{Mr_{\mathrm{ph}}-\Qcal^{2}}.
\label{eq:bcOriginal}
\end{equation}

For a static observer at $r_{\mathrm{obs}}$, the angular radius of the shadow
obeys
\begin{equation}
 \sin\vartheta_{\mathrm{sh}}
 =\frac{b_{c}\sqrt{F(r_{\mathrm{obs}})}}{r_{\mathrm{obs}}},
 \qquad
 R_{\mathrm{sh}}
 \equiv r_{\mathrm{obs}}\sin\vartheta_{\mathrm{sh}}
 =b_{c}\sqrt{F(r_{\mathrm{obs}})}.
\label{eq:shadowFinite}
\end{equation}
With the areal-radius scaling adopted in Eq.~\eqref{eq:shadowFinite},
\begin{equation}
 \lim_{r_{\mathrm{obs}}\to\infty}
 r_{\mathrm{obs}}\sin\vartheta_{\mathrm{sh}}
 =\sqrt{c}\,b_{c}.
\label{eq:shadowInfinity}
\end{equation}
This rescaled radius is convention dependent in the conical asymptotic
geometry; the directly measured invariant quantity is the angular radius
$\vartheta_{\mathrm{sh}}$.
Equations~\eqref{eq:rphOriginal}--\eqref{eq:shadowInfinity} extend the
Schwarzschild-like shadow prescription of Ref.~\cite{Pantig:2024kqy} to the
charged metric in Eq.~\eqref{eq:F}.  Reality of the circular-orbit roots
requires $9M^{2}\geq8c\Qcal^{2}$.  In the black-hole domain
\eqref{eq:domain}, this inequality is automatically satisfied.  The
physical shadow is determined by the outer unstable root lying in the
static region, with $F(r_{\mathrm{ph}})>0$, and by a static observer located
outside that orbit. The exact integral describes scattering only for $b>b_{c}$.  The post-Minkowskian series imposes the stronger conditions $\mu/b\ll1$ and $\Qbar^{2}/b^{2}\ll1$, so it should not be extrapolated into the strong-deflection regime near $b_{c}$ \cite{Bozza2002}.

\begin{figure*}[t]
 \includegraphics[width=0.485\textwidth]{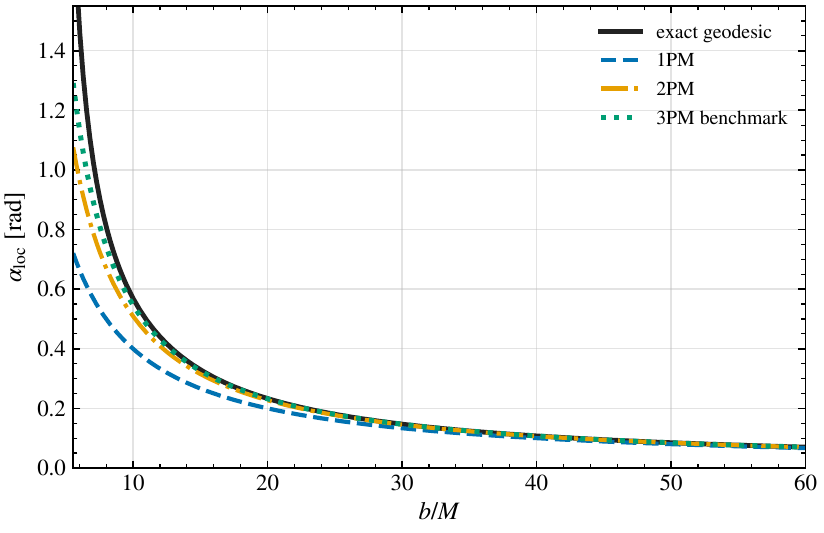}\hfill
 \includegraphics[width=0.485\textwidth]{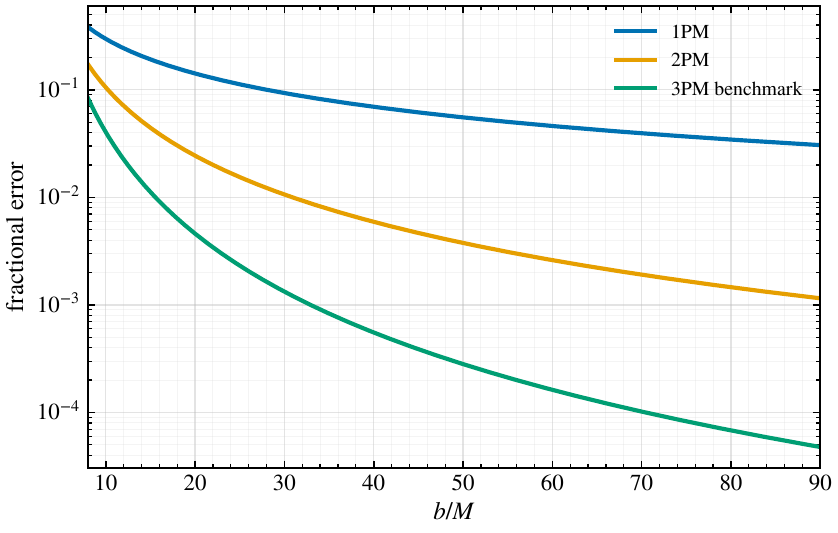}
 \caption{Vacuum method validation for the Reissner--Nordstr\"om reference values $M=1$ and $Q/M=0.6$.  Left: exact geodesic scattering compared with the 1PM, complete 2PM, and 3PM expansions.  Right: corresponding fractional errors.  The 3PM expression is a geodesic benchmark rather than part of the Gauss-Bonnet truncation.}
 \label{fig:methodValidation}
\end{figure*}

Figure~\ref{fig:methodValidation} compares the exact integral with the 1PM, 2PM, and 3PM approximations.  The visible improvement from 1PM to 2PM verifies the importance of the complete mass-squared coefficient, while the 3PM curve illustrates the expected asymptotic convergence away from the photon sphere.

\section{Weak deflection in a homogeneous plasma}
\label{sec:plasma}

\subsection{Frequency normalization and plasma optical metric}

A cold, nonmagnetized plasma introduces a dispersive scale through its electron plasma frequency $\omega_{p}$ \cite{PerlickTsupkoBisnovatyiKogan2015,BisnovatyiKoganTsupko2010}.  Because the original time coordinate is not normalized at infinity, the conserved frequency relevant for propagation is the energy associated with $T$,
\begin{equation}
 \omega_{\infty}=\mathcal{E}=\frac{E}{\sqrt{c}}.
\label{eq:frequencyNormalization}
\end{equation}
This step prevents the Lorentz-violating normalization $c$ from being counted twice in the plasma parameter.  A static observer at radius $R$ measures $\omega(R)=\omega_{\infty}/\sqrt{A(R)}$.

We define the dimensionless frequency ratio and the asymptotic group speed by
\begin{equation}
 \delta=\frac{\omega_{p}^{2}}{\omega_{\infty}^{2}},
 \qquad
 v^{2}=1-\delta,
 \qquad
 0\leq\delta<1.
\label{eq:delta}
\end{equation}
The upper bound ensures propagation in the asymptotic region.  As $\delta$ increases, the ray behaves more like a slower massive trajectory and becomes more sensitive to the gravitational potential.

For a homogeneous plasma, the refractive index is
\begin{equation}
 n^{2}(R)=1-\frac{\omega_{p}^{2}}{\omega^{2}(R)}
 =1-\delta A(R).
\label{eq:refractiveindex}
\end{equation}
The constant asymptotic value is $n^{2}(\infty)=v^{2}$.  Multiplying the vacuum optical metric by this refractive factor gives
\begin{equation}
 \dd\sigma_{p}^{2}=n^{2}(R)\left[
 \frac{\dd R^{2}}{A^{2}(R)}
 +\frac{R^{2}}{A(R)}\dd\varphi^{2}\right].
\label{eq:plasmaoptical}
\end{equation}
A constant conformal factor at infinity does not change the angular period, so the relations among $\alphaloc$, $\alphabg$, and $\alphatot$ remain those in Eq.~\eqref{eq:anglerelation}.

For a diagonal optical metric $E_{p}(R)\dd R^{2}+G_{p}(R)\dd\varphi^{2}$, the curvature density can be evaluated without separately simplifying $K_{p}$ and the area element:
\begin{equation}
 K_{p}\sqrt{E_{p}G_{p}}
 =-\frac{\dd}{\dd R}\left[
 \frac{1}{\sqrt{E_{p}}}
 \frac{\dd\sqrt{G_{p}}}{\dd R}\right].
\label{eq:plasmaKgeneral}
\end{equation}
With $N\equiv n^{2}=1-\delta A$, this identity yields the exact total derivative \cite{CrisnejoGallo2018}
\begin{equation}
 K_{p}\dd S_{p}=-\frac{\dd}{\dd R}\left[
 \sqrt{A}-\frac{RA'}{2\sqrt{A}\,N}\right]
 \dd R\dd\varphi.
\label{eq:plasmaKdSexact}
\end{equation}
This form is useful both analytically and numerically because it isolates the frequency dependence in $N$.

The 2PM expansion of the plasma curvature density is
\begin{align}
 K_{p}\dd S_{p}={}&-\frac{\mu}{R^{2}}
 \left(1+\frac{1}{v^{2}}\right)\dd R\dd\varphi
 \nonumber\\
 &+\frac{\Qbar^{2}}{R^{3}}
 \left(1+\frac{2}{v^{2}}\right)\dd R\dd\varphi
 \nonumber\\
 &+\frac{\mu^{2}}{R^{3}}
 \left(-1-\frac{6}{v^{2}}+\frac{4}{v^{4}}\right)
 \dd R\dd\varphi+\Order(\epsilon^{3}).
\label{eq:plasmaKdSexp}
\end{align}
The frequency dependence is not a common multiplicative factor.  Consequently, a plasma changes the relative weights of the leading mass, 2PM mass, and charge sectors rather than simply rescaling the vacuum angle.

\subsection{Gauss-Bonnet result and physical enhancement factors}

The asymptotic trajectory has geometric impact parameter
\begin{equation}
 b=\frac{J}{\mathcal{E}v}.
\label{eq:plasmab}
\end{equation}
The factor of $v$ distinguishes the geometric distance of the asymptotic line from the ratio $J/\mathcal{E}$ used for a vacuum photon.  In terms of the conserved quantities defined in Eq.~\eqref{eq:bInvariant},
\begin{equation}
 \frac{L}{E}=\frac{J}{\mathcal{E}}=vb
 \qquad\text{(homogeneous plasma)}.
\label{eq:plasmaImpactDictionary}
\end{equation}
Thus the symbol $b$ denotes the geometric impact parameter in both media, but it equals $L/E$ only in vacuum.

The first corrected plasma orbit is
\begin{equation}
 U_{\gamma}^{p}(\varphi)=\frac{\sin\varphi}{b}
 +\frac{\mu}{b^{2}}
 \left(\frac{1}{v^{2}}+\cos^{2}\varphi\right)
 +\Order(\epsilon^{2}).
\label{eq:plasmaorbitfirst}
\end{equation}
Relative to the vacuum boundary in Eq.~\eqref{eq:orbitfirstvac}, the constant part is enhanced by $v^{-2}$, reflecting the longer interaction time of a slower dispersive ray.

For finite source and receiver radii, the endpoint angles also require the
leading charge deformation of the orbit.  Retaining terms linear in $\mu$ and
linear in $\Qbar^{2}$ gives
\begin{widetext}
\begin{align}
 U_{\gamma}^{p}(\varphi)={}&\frac{\sin\varphi}{b}
 +\frac{\mu}{b^{2}}
 \left(\frac{1}{v^{2}}+\cos^{2}\varphi\right)
 +\frac{\Qbar^{2}}{b^{3}}
 \left[
 \frac{2+v^{2}}{4v^{2}}
 \left(\varphi-\frac{\pi}{2}\right)\cos\varphi
 -\frac{\sin3\varphi}{16}
 -\frac{8+v^{2}}{16v^{2}}\sin\varphi
 \right]+\Order\left(\frac{\mu^{2}}{b^{3}},
 \frac{\mu\Qbar^{2}}{b^{4}},
 \frac{\Qbar^{4}}{b^{5}}\right).
\label{eq:plasmaOrbitFinite}
\end{align}
\end{widetext}
The homogeneous term in the charge sector is fixed by requiring
Eq.~\eqref{eq:plasmaOrbitFinite} to satisfy both the Binet equation and the
first integral in Eq.~\eqref{eq:plasmaorbitexact}.  This requirement is
important for the finite-distance endpoint contribution.

For a symmetric finite-distance configuration with
$R_{S}=R_{R}=R$, define
\begin{equation}
 x=\frac{b}{R}=\frac{\sqrt{c}\,b}{r},
 \qquad
 D=1-x^{2}.
\label{eq:finiteXD}
\end{equation}
Here $0<x<1$.  Because the endpoint expansion is nonuniform as
$D\to0$, a sufficient separation from the tangential-endpoint limit is
$\mu/(bv^{2})\ll D$ and $\Qbar^{2}/(b^{2}v^{2})\ll D$.
The local finite-distance weak angle then becomes
\begin{widetext}
\begin{eqnarray}
 \alpha_{\mathrm{loc}}^{p}(R)
 =\frac{2\mu(1+v^{2})}{bv^{2}}\sqrt{D}
 -\frac{\Qbar^{2}(2+v^{2})}{2b^{2}v^{2}}
 \left[\arccos x+x\sqrt{D}\right] \notag
\Order\left(\frac{\mu^{2}}{b^{2}},
 \frac{\mu\Qbar^{2}}{b^{3}},
 \frac{\Qbar^{4}}{b^{4}}\right).
\label{eq:finiteLocalPlasma}
\end{eqnarray}
\end{widetext}
The normalized source-receiver angular separation is
\begin{equation}
 \varphi_{RS}
 =\pi-2\arcsin x
 +\frac{2\mu}{b}\frac{v^{-2}+D}{\sqrt{D}}
 +\frac{2\Qbar^{2}}{b^{2}}
 \frac{\mathcal{A}_{Q}(x,v)}{\sqrt{D}},
\label{eq:finiteVarphiRS}
\end{equation}
where
\begin{equation}
 \mathcal{A}_{Q}(x,v)
 =-\frac{2+v^{2}}{4v^{2}}\arccos x\sqrt{D}
 -\frac{\sin\!\left(3\arcsin x\right)}{16}
 -\frac{8+v^{2}}{16v^{2}}x.
\label{eq:AQfinite}
\end{equation}
Using Eq.~\eqref{eq:finiteAngleConversion}, the weak angle in the original
azimuth is therefore
\begin{equation}
 \alpha_{\phi}^{p}(r)
 =\alpha_{\mathrm{loc}}^{p}(R)
 +\left(c^{-1/2}-1\right)\varphi_{RS}.
\label{eq:finiteCoordinatePlasma}
\end{equation}
No expansion in $c-1$ is needed in
Eqs.~\eqref{eq:finiteLocalPlasma}--\eqref{eq:finiteCoordinatePlasma}.

Integrating Eq.~\eqref{eq:plasmaKdSexp} over the domain bounded by Eq.~\eqref{eq:plasmaorbitfirst} gives
\begin{equation}
 \begin{aligned}
 \alpha_{\mathrm{loc}}^{p}={}&\frac{2\mu}{b}
 \left(1+\frac{1}{v^{2}}\right)
 +\frac{3\pi\mu^{2}}{4b^{2}}
 \left(1+\frac{4}{v^{2}}\right)\\
 &-\frac{\pi\Qbar^{2}}{4b^{2}}
 \left(1+\frac{2}{v^{2}}\right)
 +\Order(\epsilon^{3}).
 \end{aligned}
\label{eq:plasmaalpha}
\end{equation}
The vacuum expression follows continuously as $v\rightarrow1$.  For $0<v<1$, the mass attraction and the magnitude of the negative charge correction both increase, but by different amounts.

The corresponding coordinate angular excess follows from the same conical
boundary term,
\begin{equation}
\begin{aligned}
 \alpha_{\phi}^{p}={}&\pi(c^{-1/2}-1)
 +\frac{2M}{c^{2}b}\left(1+\frac{1}{v^{2}}\right)\\
 &+\frac{3\pi M^{2}}{4c^{7/2}b^{2}}
 \left(1+\frac{4}{v^{2}}\right)
 -\frac{\pi\Qcal^{2}}{4c^{5/2}b^{2}}
 \left(1+\frac{2}{v^{2}}\right)
 +\Order(\epsilon^{3}).
\end{aligned}
\label{eq:plasmaCoordinateAngle}
\end{equation}
The constant term remains achromatic; all chromaticity resides in the
background-subtracted curvature contribution.

The far-distance limit of Eq.~\eqref{eq:finiteCoordinatePlasma} reproduces
the leading mass and charge sectors of Eq.~\eqref{eq:plasmaCoordinateAngle}.
Keeping the complete mass contribution already obtained at 2PM order, and
substituting Eq.~\eqref{eq:cQ}, we can write the final coordinate angle directly
in terms of the two Lorentz-violating couplings as
\begin{align}
 \alpha_{\phi}^{p}={}&
 \pi\left[
 \sqrt{\frac{1+l_{1}-l_{2}/2}{1+l_{1}}}-1
 \right]
 \nonumber\\
 &+\frac{2M}{b}\left(1+\frac{1}{v^{2}}\right)
 \left(\frac{1+l_{1}-l_{2}/2}{1+l_{1}}\right)^{2}
 \nonumber\\
 &+\frac{3\pi M^{2}}{4b^{2}}
 \left(1+\frac{4}{v^{2}}\right)
 \left(\frac{1+l_{1}-l_{2}/2}{1+l_{1}}\right)^{7/2}
 \nonumber\\
 &-\frac{\pi(1-l_{1})Q^{2}}{4b^{2}(1-l_{1}-l_{2}/2)^{2}}
 \left(1+\frac{2}{v^{2}}\right)
 \left(\frac{1+l_{1}-l_{2}/2}{1+l_{1}}\right)^{5/2}
 +\Order(\epsilon^{3}).
\label{eq:plasmaCoordinateCouplings}
\end{align}
In the vacuum limit $\delta\to0$ (and hence $v\to1$),
Eq.~\eqref{eq:plasmaCoordinateCouplings} reduces to
\begin{align}
 \alpha_{\phi}^{\rm photon}={}&
 \pi\left[
 \sqrt{\frac{1+l_{1}-l_{2}/2}{1+l_{1}}}-1
 \right]
 +\frac{4M}{b}
 \left(\frac{1+l_{1}-l_{2}/2}{1+l_{1}}\right)^{2}
 \nonumber\\
 &+\frac{15\pi M^{2}}{4b^{2}}
 \left(\frac{1+l_{1}-l_{2}/2}{1+l_{1}}\right)^{7/2}
 \nonumber\\
 &-\frac{3\pi(1-l_{1})Q^{2}}
 {4b^{2}(1-l_{1}-l_{2}/2)^{2}}
 \left(\frac{1+l_{1}-l_{2}/2}{1+l_{1}}\right)^{5/2}
 +\Order(\epsilon^{3}).
\label{eq:photonCoordinateCouplings}
\end{align}
The first-order coupling expansion makes the separation of the two directions
particularly transparent.  We obtain
\begin{align}
 \alpha_{\phi}^{p}={}&
 \frac{2M}{b}\left(1+\frac{1}{v^{2}}\right)
 +\frac{3\pi M^{2}}{4b^{2}}
 \left(1+\frac{4}{v^{2}}\right)
 -\frac{\pi Q^{2}}{4b^{2}}
 \left(1+\frac{2}{v^{2}}\right)
 \nonumber\\
 &-\frac{\pi l_{2}}{4}
 -\frac{2Ml_{2}}{b}\left(1+\frac{1}{v^{2}}\right)
 -\frac{21\pi M^{2}l_{2}}{16b^{2}}
 \left(1+\frac{4}{v^{2}}\right)
 \nonumber\\
 &-\frac{\pi Q^{2}}{4b^{2}}
 \left(1+\frac{2}{v^{2}}\right)
 \left(l_{1}-\frac{l_{2}}{4}\right)
 +\Order(l_{i}^{2},\epsilon^{3}).
\label{eq:plasmaCoordinateLinearCouplings}
\end{align}
In the same vacuum-photon limit this becomes
\begin{align}
 \alpha_{\phi}^{\rm photon}={}&
 \frac{4M}{b}
 +\frac{15\pi M^{2}}{4b^{2}}
 -\frac{3\pi Q^{2}}{4b^{2}}
 -\frac{\pi l_{2}}{4}
 -\frac{4Ml_{2}}{b}
 \nonumber\\
 &-\frac{105\pi M^{2}l_{2}}{16b^{2}}
 -\frac{3\pi Q^{2}}{4b^{2}}
 \left(l_{1}-\frac{l_{2}}{4}\right)
 +\Order(l_{i}^{2},\epsilon^{3}).
\label{eq:photonCoordinateLinearCouplings}
\end{align}
The constant term is governed by $l_{2}$ at linear order.  The same coupling
also changes the mass sector, while $l_{1}$ first enters through the charged
$b^{-2}$ contribution.  Setting $l_{1}=l_{2}=0$ recovers the standard
Reissner--Nordstr\"om weak angle.

Dividing each plasma coefficient by its vacuum counterpart isolates three dimensionless enhancement factors:
\begin{equation}
\begin{aligned}
 \mathcal{R}_{M}^{(1)}&=\frac{1+v^{-2}}{2},
 &\mathcal{R}_{M}^{(2)}&=\frac{1+4v^{-2}}{5},\\
 \mathcal{R}_{Q}^{(2)}&=\frac{1+2v^{-2}}{3}.
\end{aligned}
\label{eq:enhancementRatios}
\end{equation}
These unequal functions provide a chromatic discriminator.  A multi-frequency data set would not merely measure a larger angle at lower frequency; it would change the relative importance of the $b^{-1}$ and $b^{-2}$ terms.

At leading order, the local thin-lens Einstein radius in a homogeneous plasma becomes
\begin{equation}
 \theta_{E}^{p}=\left[
 \frac{2\mu D_{LS}}{D_{L}D_{S}}
 \left(1+\frac{1}{v^{2}}\right)\right]^{1/2}
 =\theta_{E}^{\rm vac}
 \left(\frac{1+v^{-2}}{2}\right)^{1/2}.
\label{eq:EinsteinPlasma}
\end{equation}
This relation is a transparent observable consequence of the leading term, but it is only an analytic baseline.  A realistic radial density profile adds refractive-gradient bending and changes both the lens equation and the frequency dependence.

\subsection{Hamiltonian check and exact plasma integral}

The plasma ray can also be derived from the Hamiltonian
\begin{equation}
 \mathcal{H}=\frac{1}{2}\left(g^{\mu\nu}p_{\mu}p_{\nu}
 +\omega_{p}^{2}\right)=0.
\label{eq:plasmaHamiltonian}
\end{equation}
This formulation treats the photon in a homogeneous plasma as an effective massive excitation and supplies an independent check of the optical-metric result.

The exact orbit equation is
\begin{equation}
 \left(\frac{\dd U}{\dd\varphi}\right)^{2}
 =\frac{1-\delta A(U)}{v^{2}b^{2}}-A(U)U^{2},
\label{eq:plasmaorbitexact}
\end{equation}
where $A(U)=1-2\mu U+\Qbar^{2}U^{2}$.  Differentiating gives
\begin{equation}
 U''+U=\mu a+3\mu U^{2}-\Qbar^{2}aU-2\Qbar^{2}U^{3},
 \qquad
 a=\frac{1-v^{2}}{v^{2}b^{2}}.
\label{eq:plasmaBinet}
\end{equation}
The first iterate of Eq.~\eqref{eq:plasmaBinet} is precisely Eq.~\eqref{eq:plasmaorbitfirst}, so the perturbed boundary used by the Gauss-Bonnet method is dynamically consistent.

If $U_{m}^{p}$ is the first positive root of the right-hand side of Eq.~\eqref{eq:plasmaorbitexact}, the exact local plasma angle is
\begin{equation}
 \alpha_{\mathrm{loc}}^{p,\mathrm{exact}}
 =2\int_{0}^{U_{m}^{p}}
 \frac{\dd U}
 {\sqrt{[1-\delta A(U)]/(v^{2}b^{2})-A(U)U^{2}}}-\pi.
\label{eq:plasmaexactintegral}
\end{equation}
Expanding this integral reproduces Eq.~\eqref{eq:plasmaalpha}.  The agreement checks the frequency normalization, the definition of $b$, and the plasma curvature density simultaneously.

The capture threshold is itself frequency dependent.  The accessible outer
unstable circular plasma orbit at $R=R_{c,p}$ satisfies
\begin{equation}
 2A(R_{c,p})\left[1-\delta A(R_{c,p})\right]
 -R_{c,p}A'(R_{c,p})=0,
\label{eq:plasmaCriticalRadius}
\end{equation}
and its geometric critical impact parameter is
\begin{equation}
 b_{c,p}^{2}=\frac{R_{c,p}^{2}
 [1-\delta A(R_{c,p})]}{A(R_{c,p})v^{2}}.
\label{eq:plasmaCriticalImpact}
\end{equation}
A scattering trajectory requires $b>b_{c,p}$ and
$N(R)=1-\delta A(R)>0$ along the complete path.  Equations
\eqref{eq:plasmaCriticalRadius} and \eqref{eq:plasmaCriticalImpact} reduce
to the vacuum photon-sphere conditions as $v\rightarrow1$.

The weak-field condition becomes increasingly restrictive as the asymptotic frequency approaches the plasma cutoff.  In addition to $\Qbar^{2}/b^{2}\ll1$, a useful sufficient condition is
\begin{equation}
 \frac{\mu}{bv^{2}}\ll1.
\label{eq:plasmaValidity}
\end{equation}
The apparent divergence of Eq.~\eqref{eq:plasmaalpha} as $v\rightarrow0$ is therefore outside the uniform domain of the expansion and must not be interpreted as a physical infinite deflection.

\begin{figure*}[t]
 \includegraphics[width=0.485\textwidth]{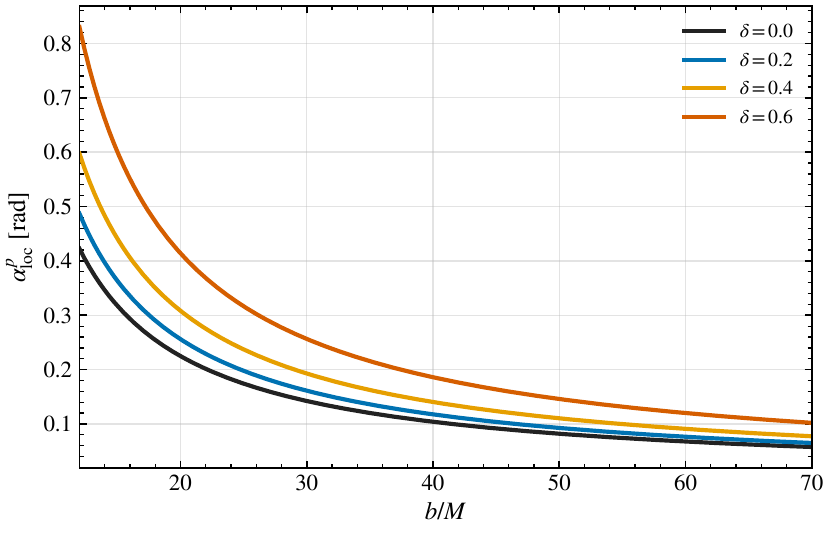}\hfill
 \includegraphics[width=0.485\textwidth]{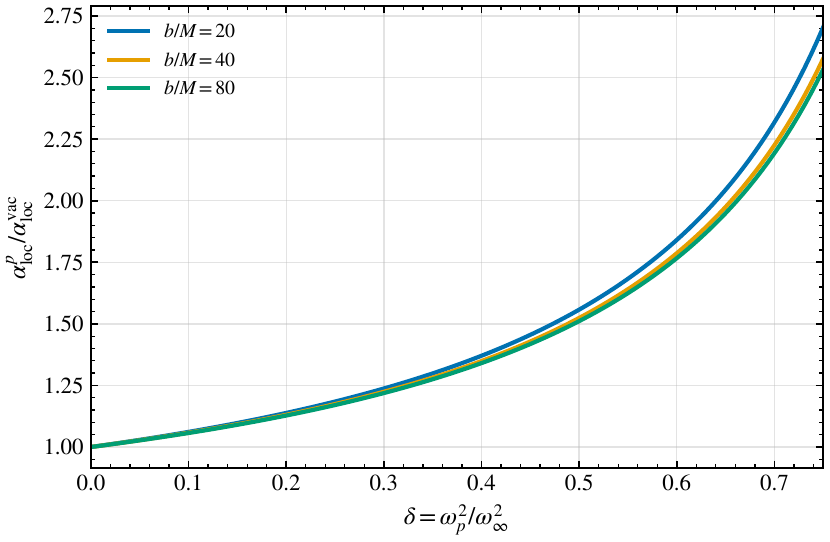}
 \caption{Homogeneous-plasma signatures for $M=1$, $Q/M=0.6$, and $l_{1}=l_{2}=0.04$.  Left: exact local deflection for several values of $\delta=\omega_{p}^{2}/\omega_{\infty}^{2}$.  Right: exact ratio $\alpha_{\rm loc}^{p}/\alpha_{\rm loc}^{\rm vac}$ as a function of $\delta$ for three impact parameters.  At fixed $b$, the exact scattering branch terminates at $b=b_{c,p}$ before $\delta=1$; the displayed range also should be read together with the 2PM validity condition in Eq.~\eqref{eq:plasmaValidity}.}
 \label{fig:plasmaPhysics}
\end{figure*}

Figure~\ref{fig:plasmaPhysics} shows both the absolute chromatic increase and the dimensionless enhancement ratio.  The weak dependence of the ratio on $b$ at large impact parameter reflects dominance of the leading mass term.

\section{Numerical validation and coupling response}
\label{sec:numerics}

\subsection{Exact quadrature and benchmark choices}

The exact vacuum and plasma integrals contain an integrable square-root singularity at the turning point.  We remove it by setting
\begin{equation}
 U=U_{m}(1-s^{2}),
 \qquad 0\leq s\leq1,
\label{eq:endpointSubstitution}
\end{equation}
which maps the singular endpoint to a finite integrand.  The numerical routine selects the first positive root and rejects parameter sets without a scattering branch.  In vacuum it verifies the horizon and photon-sphere conditions in Eq.~\eqref{eq:bc}; in plasma it instead checks $N>0$, $b>b_{c,p}$, and the frequency-dependent critical-orbit conditions in Eqs.~\eqref{eq:plasmaCriticalRadius} and \eqref{eq:plasmaCriticalImpact}.

Unless stated otherwise, the plots use $M=1$ and $Q/M=0.6$.  Couplings of order $10^{-1}$ are used only as sensitivity benchmarks that make the functional dependence visible on ordinary plot scales.  They are not interpreted as observationally allowed values.  For small couplings, the curves scale according to Eqs.~\eqref{eq:cLinear}-\eqref{eq:linearalpha}.

Table~\ref{tab:validation} reports representative exact and 2PM values.  The error decreases with $b$, as expected for an asymptotic expansion, and is larger in plasma at the same $b$ because the effective expansion parameter contains $v^{-2}$.
\begin{table}[t]
 \caption{Exact and 2PM local deflection angles for $l_{1}=l_{2}=0.04$ and $Q/M=0.6$.  Angles are in radians.}
 \label{tab:validation}
 \begin{ruledtabular}
 \begin{tabular}{ccccc}
 $b/M$ & $\delta$ & $\alpha_{\rm exact}$ & $\alpha_{\rm 2PM}$ & error \\
 \hline
 12 & 0   & 0.42250 & 0.39479 & $6.56\%$ \\
 12 & 0.4 & 0.59750 & 0.54114 & $9.43\%$ \\
 20 & 0   & 0.22491 & 0.21983 & $2.26\%$ \\
 20 & 0.4 & 0.30824 & 0.29842 & $3.19\%$ \\
 50 & 0   & 0.08208 & 0.08179 & $0.35\%$ \\
 50 & 0.4 & 0.11045 & 0.10991 & $0.49\%$ \\
 \end{tabular}
 \end{ruledtabular}
\end{table}

\begin{figure}[t]
 \includegraphics[width=\columnwidth]{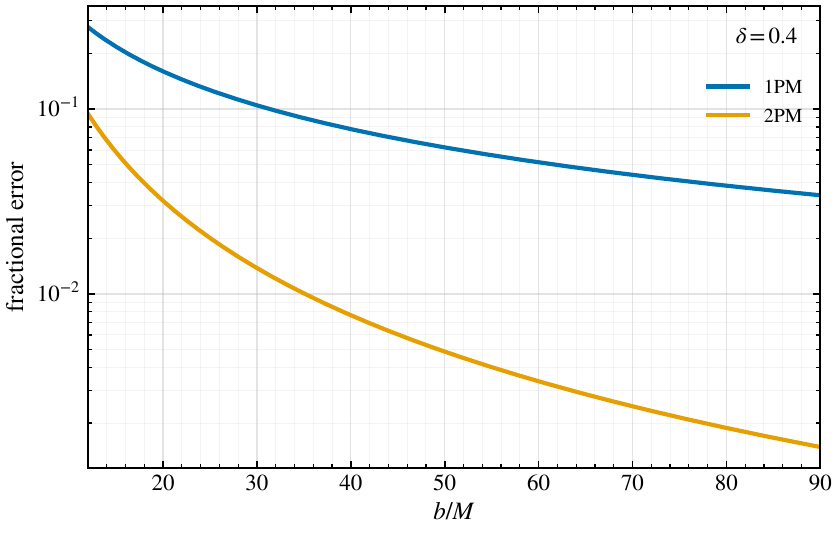}
 \caption{Fractional errors of the 1PM and 2PM homogeneous-plasma formulas relative to Eq.~\eqref{eq:plasmaexactintegral} for $\delta=0.4$, $l_{1}=l_{2}=0.04$, $M=1$, and $Q/M=0.6$.}
 \label{fig:plasmaError}
\end{figure}

The plasma truncation error is displayed in Fig.~\ref{fig:plasmaError}.  At fixed $\delta$, the 2PM curve approaches the exact result monotonically over the weak-field interval shown.

\subsection{Separating the two coupling directions}

\begin{figure*}[t]
 \includegraphics[width=0.485\textwidth]{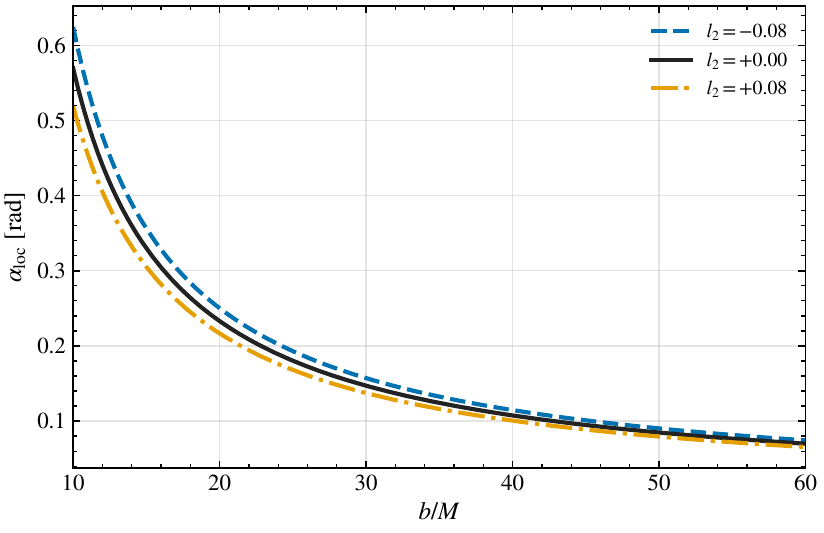}\hfill
 \includegraphics[width=0.485\textwidth]{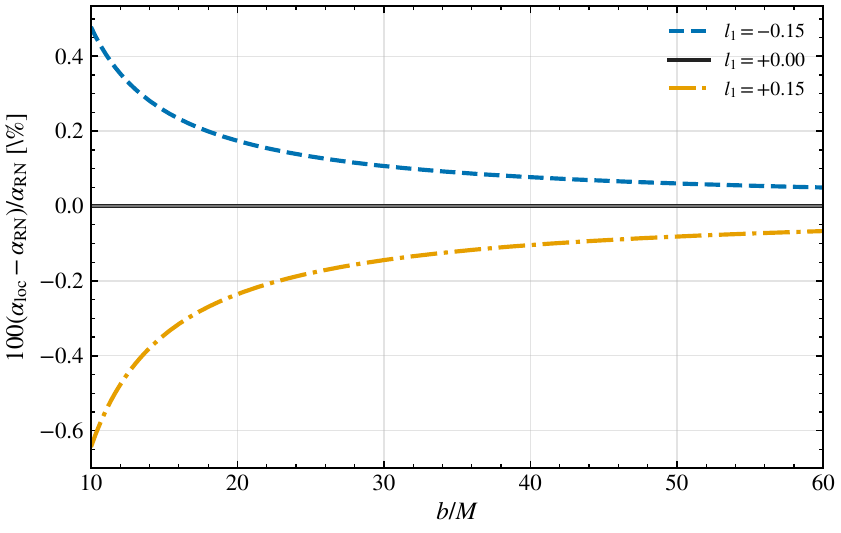}
 \caption{Exact vacuum scattering from Eq.~\eqref{eq:exactvacintegral}.  Left: local angle for three values of $l_{2}$ at fixed $l_{1}=0$.  Right: percentage shift relative to Reissner--Nordstr\"om for three values of $l_{1}$ at fixed $l_{2}=0$.  We use $M=1$ and $Q/M=0.6$.  The different vertical scales are the numerical manifestation of Eqs.~\eqref{eq:muLinear} and \eqref{eq:qLinear}.}
 \label{fig:couplings}
\end{figure*}

Figure~\ref{fig:couplings} varies one coupling at a time.  Changing $l_{2}$ shifts the dominant mass normalization and therefore changes the full angle already at order $b^{-1}$.  Changing $l_{1}$ affects the charge sector, so its relative signature is weaker and decays more rapidly with impact parameter.

\begin{figure}[t]
 \includegraphics[width=\columnwidth]{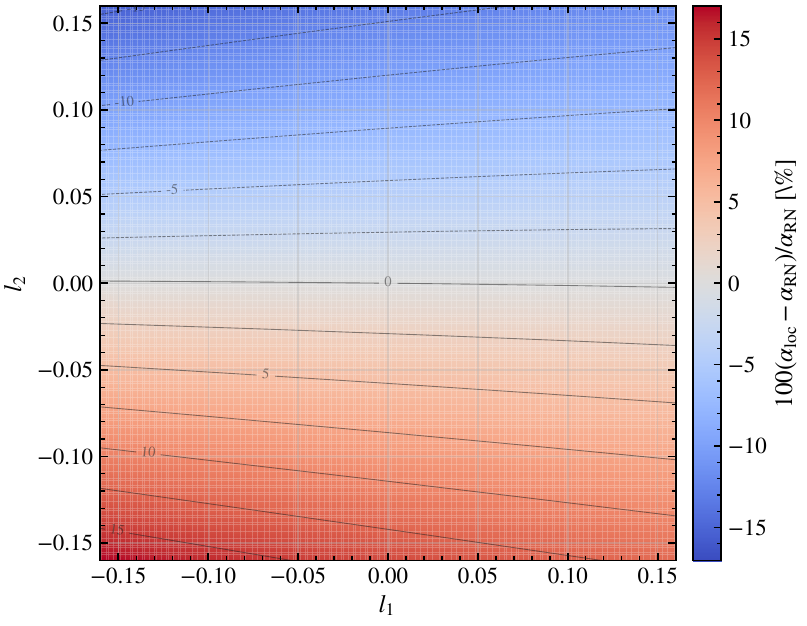}
 \caption{Percentage change of the 2PM local vacuum angle relative to Reissner--Nordstr\"om over the $(l_{1},l_{2})$ plane at $b/M=20$ and $Q/M=0.6$.  Every displayed point satisfies the black-hole conditions in Eq.~\eqref{eq:domain}.}
 \label{fig:couplingMap}
\end{figure}

The combined finite-coupling response is shown in Fig.~\ref{fig:couplingMap}.  The near-horizontal contours demonstrate that $l_{2}$ dominates the local angle at the chosen impact parameter, while $l_{1}$ modulates the smaller charge contribution.

A small-coupling angular scale helps distinguish mathematical normalization from an observable displacement.  For the illustrative value $l_{2}=10^{-10}$, the conical term is
\begin{equation}
\begin{aligned}
 \left|\pi(c^{-1/2}-1)\right|
 &=\frac{\pi|l_{2}|}{4}+\Order(l_{i}^{2})\\
 &\simeq7.85\times10^{-11}\ {\rm rad}
 \simeq16.2\ \mu{\rm as}.
\end{aligned}
\label{eq:conicalScale}
\end{equation}
This number is not a predicted image shift and is not a bound.  It is the scale of the coordinate angular identification; converting it into image positions requires a global conical lens geometry, finite source and observer distances, and a consistent mass calibration.

\section{Steady spherical matter accretion}
\label{sec:accretion}

We now consider the steady radial accretion of a neutral test fluid onto the
black-hole geometry in Eq.~\eqref{eq:originalmetric}.  
The fluid is assumed to be non-self-gravitating, inviscid, and adiabatic,
so that its backreaction on the metric is neglected.

The
relativistic Bondi-Michel formalism and its Hamiltonian formulation for a
general static, spherically symmetric metric are reviewed in
Refs.~\cite{Bondi1952Accretion,Michel1972Accretion,
AhmedEtAl2016Accretion,AzregAinou2017Accretion}.

The stress-energy tensor of a perfect fluid is
\begin{equation}
 T^{\mu\nu}=(\rho+p)u^{\mu}u^{\nu}+p\,g^{\mu\nu},
\label{eq:accStressTensor}
\end{equation}
where $\rho$ is the total energy density in the fluid rest frame, $p$ is the
pressure, and $u^{\mu}$ is the future-directed four-velocity.  The baryon
number density is denoted by $n$; it must be distinguished from $\rho$ unless
the fluid is exactly pressureless, in which case $\rho=m_{b}n$ with $m_b$ the
rest mass per baryon.  Stationarity and spherical symmetry imply
\begin{equation}
 u^{\mu}=(u^{t},u,0,0),
 \qquad u\equiv u^{r},
\label{eq:accFourVelocity}
\end{equation}
with $u<0$ for accretion and $u>0$ for an outflow.  The normalization
$u^{\mu}u_{\mu}=-1$ gives
\begin{align}
 -F(u^{t})^{2}+\frac{u^{2}}{F}&=-1,\nonumber\\
 u^{t}&=\frac{\sqrt{F+u^{2}}}{F},
 &u_{t}&=-\sqrt{F+u^{2}}.
\label{eq:accNormalization}
\end{align}

For a neutral fluid, the relevant conservation equations are
\begin{equation}
 \nabla_{\mu}(n u^{\mu})=0,
 \qquad
 \nabla_{\mu}T^{\mu\nu}=0.
\label{eq:accConservation}
\end{equation}
Since $\sqrt{-g}=r^{2}\sin\theta$, particle-number conservation integrates
immediately to
\begin{equation}
 r^{2}n u=\mathcal C_{N},
 \qquad
 \dot{\mathcal N}\equiv-4\pi\mathcal C_{N}>0
 \quad\hbox{for accretion}.
\label{eq:accParticleFlux}
\end{equation}
Here $\dot{\mathcal N}$ is the positive baryon accretion rate per unit
coordinate time $t$, and $\mathcal C_N<0$ on the inflowing branch.  The
corresponding rest-mass accretion rate per unit $t$ is
$\dot M_{0}=m_b\dot{\mathcal N}$.

Introduce the specific enthalpy
\begin{equation}
 h\equiv\frac{\rho+p}{n}.
\label{eq:accEnthalpy}
\end{equation}
For an isentropic flow the first law gives
\begin{equation}
 \dd\rho=h\,\dd n,
 \qquad
 \dd p=n\,\dd h.
\label{eq:accThermodynamics}
\end{equation}
The timelike Killing vector $\xi^{\mu}=(\partial_t)^{\mu}$ yields a second
conserved flux,
\begin{equation}
 r^{2}T^{r}{}_{t}
 =r^{2}(\rho+p)u u_{t}
 =\mathcal C_{E}.
\label{eq:accEnergyFlux}
\end{equation}
Dividing Eq.~\eqref{eq:accEnergyFlux} by
Eq.~\eqref{eq:accParticleFlux} gives the relativistic Bernoulli integral
\begin{align}
 \mathcal B&\equiv-hu_{t}=h\sqrt{F+u^{2}}
 =\text{constant},\nonumber\\
 \mathcal C_{E}&=-\mathcal C_{N}\mathcal B.
\label{eq:accBernoulli}
\end{align}
Consequently, the positive Killing-energy accretion rate is
$\dot E=4\pi\mathcal C_E=\dot{\mathcal N}\mathcal B$.
Because $F(r)\to c$ rather than unity, $\partial_t$ is not unit normalized at
infinity.  The Bernoulli constant and rates referred to the unit-normalized
asymptotic time translation $\widehat\xi=\xi/\sqrt c$ and to the
corresponding time $T=\sqrt c\,t$ are therefore
\begin{equation}
 \widehat{\mathcal B}=\frac{\mathcal B}{\sqrt c},
 \qquad
 \dot{\mathcal N}_{\infty}=\frac{\dot{\mathcal N}}{\sqrt c},
 \qquad
 \dot E_{\infty}=\frac{\dot E}{c}
 =\widehat{\mathcal B}\,\dot{\mathcal N}_{\infty}.
\label{eq:accPhysicalEnergy}
\end{equation}
This normalization is essential when accretion rates are compared between
different values of $l_1$ and $l_2$.

The signed radial three-velocity measured by a static orthonormal observer is
\begin{equation}
 v\equiv\frac{u}{\sqrt{F+u^{2}}},
 \qquad
 u^{2}=\frac{Fv^{2}}{1-v^{2}},
 \qquad |v|<1,
\label{eq:accLocalVelocity}
\end{equation}
where $v<0$ denotes inflow.  Equations~\eqref{eq:accParticleFlux} and
\eqref{eq:accLocalVelocity} imply
\begin{equation}
 n(r,v)=\frac{|\mathcal C_N|}{r^{2}}
 \left(\frac{1-v^{2}}{Fv^{2}}\right)^{1/2}.
\label{eq:accDensityRV}
\end{equation}
The square of the Bernoulli constant may then be used as a Hamiltonian on the
$(r,v)$ phase plane,
\begin{equation}
 \mathcal H(r,v)=h[n(r,v)]^{2}\frac{F(r)}{1-v^{2}}
 =\mathcal B^{2}.
\label{eq:accHamiltonian}
\end{equation}
This form is useful because it keeps the local speed bounded and makes clear
that an equation of state is required to determine $h(n)$ and hence a unique
flow.

For an isentropic barotrope, define the sound speed by
\begin{equation}
 a^{2}\equiv\left(\frac{\partial p}{\partial\rho}\right)_{s}
 =\frac{\dd\ln h}{\dd\ln n},
 \qquad 0\leq a^{2}\leq1.
\label{eq:accSoundSpeed}
\end{equation}
Logarithmically differentiating the particle and Bernoulli integrals gives
the exact wind equation
\begin{equation}
 \frac{\dd u}{\dd r}
 =\frac{u\left[4a^{2}(F+u^{2})-rF'\right]}
 {2r\left[u^{2}-a^{2}(F+u^{2})\right]},
\label{eq:accWindEquation}
\end{equation}
where $F'=\dd F/\dd r$.  A smooth transonic solution must pass through a
critical point at which the numerator and denominator vanish simultaneously:
\begin{align}
 u_{c}^{2}&=\frac{r_{c}F'(r_c)}{4},
\label{eq:accCriticalU}\\
 v_{c}^{2}=a_{c}^{2}
 &=\frac{r_{c}F'(r_c)}{r_{c}F'(r_c)+4F(r_c)}.
\label{eq:accCriticalA}
\end{align}
For the black-hole potential in Eq.~\eqref{eq:F}, these conditions
become
\begin{align}
 u_c^{2}&=\frac{Mr_c-\Qcal^{2}}{2r_c^{2}},
\label{eq:accCriticalExplicitU}\\
 a_c^{2}&=\frac{Mr_c-\Qcal^{2}}
 {2cr_c^{2}-3Mr_c+\Qcal^{2}}.
\label{eq:accCriticalExplicitA}
\end{align}
The critical radius is therefore modified both by the asymptotic factor $c$
and by the effective charge $\Qcal$.

The same result takes a particularly transparent form in the locally
normalized variables of Eqs.~\eqref{eq:TR}-\eqref{eq:effectiveparameters}.
Writing $u^{R}=u/\sqrt c$ and $R=r/\sqrt c$, one obtains
\begin{equation}
 (u_c^{R})^{2}=\frac{R_c A'(R_c)}{4},
 \qquad
 a_c^{2}=\frac{R_cA'(R_c)}{R_cA'(R_c)+4A(R_c)}.
\label{eq:accNormalizedCritical}
\end{equation}
Thus, at fixed $(\mu,\Qbar)$, the local sonic structure is exactly that of a
Reissner--Nordstr\"om flow.  The conical factor remains in the physical area
$4\pi cR^{2}$ and hence in the integrated baryon rate
$\dot{\mathcal N}_{\infty}=-4\pi cR^{2}nu^{R}$.  This separates the local
effective-potential effect from the global solid-angle normalization.

As a simple analytic example, a constant-sound-speed equation of state
$p=k\rho$ (often called isothermal in the accretion literature) has
$a^{2}=k$ and
\begin{equation}
 \rho=K n^{1+k},
 \qquad
 h=(1+k)K n^{k},
 \qquad 0<k<1,
\label{eq:accIsothermalEOS}
\end{equation}
where $K>0$ is a constant.  Up to an irrelevant positive normalization, the
Hamiltonian reduces to
\begin{equation}
 \widetilde{\mathcal H}_{k}(r,v)
 =\frac{F(r)^{1-k}}
 {r^{4k}(v^{2})^{k}(1-v^{2})^{1-k}}.
\label{eq:accIsothermalHamiltonian}
\end{equation}
The outer critical radius is
\begin{equation}
 r_c=\frac{M(1+3k)
 +\sqrt{M^{2}(1+3k)^{2}-8ck(1+k)\Qcal^{2}}}
 {4ck},
\label{eq:accIsothermalCriticalRadius}
\end{equation}
provided the discriminant is nonnegative and the selected root lies outside
$r_+$.  In the normalized variables this becomes
\begin{equation}
 R_c=\frac{\mu(1+3k)
 +\sqrt{\mu^{2}(1+3k)^{2}-8k(1+k)\Qbar^{2}}}
 {4k}.
\label{eq:accNormalizedIsothermalCriticalRadius}
\end{equation}
If the gas is asymptotically at rest, the Bernoulli condition gives
\begin{equation}
 \frac{n_c}{n_\infty}
 =\left[\frac{c(1-k)}{F(r_c)}\right]^{1/(2k)}
 =\left[\frac{1-k}{A(R_c)}\right]^{1/(2k)},
\label{eq:accIsothermalCriticalDensity}
\end{equation}
and the regular transonic branch has
\begin{equation}
 \dot{\mathcal N}_{\infty}
 =\frac{4\pi r_c^{2}n_c|u_c|}{\sqrt c}
 =4\pi cR_c^{2}n_c|u_c^{R}|.
\label{eq:accIsothermalRate}
\end{equation}
Thus the sonic regularity condition and the asymptotic thermodynamic data
select the flux; the two integration constants cannot be assigned
independently.

Alternatively, for a relativistic polytrope one may take
\begin{align}
 p&=K n^{\gamma},
 &\rho&=m_b n+\frac{p}{\gamma-1},\nonumber\\
 h&=m_b+\frac{\gamma K}{\gamma-1}n^{\gamma-1},
 &a^{2}&=\frac{\gamma p}{\rho+p}.
\label{eq:accPolytropicEOS}
\end{align}
In that case Eqs.~\eqref{eq:accBernoulli}, \eqref{eq:accParticleFlux}, and
\eqref{eq:accCriticalU}-\eqref{eq:accCriticalA}, together with the boundary
values of $n$ and $a$, determine the transonic accretion rate.  Thus the
inequality $p\ll\rho$ alone is not sufficient to determine $u(r)$ and
$\rho(r)$; an equation of state and boundary data are indispensable.

For a transparent analytic benchmark, consider pressureless dust,
\begin{equation}
 p=0,
 \qquad
 \rho=m_b n,
 \qquad
 h=m_b.
\label{eq:accDustEOS}
\end{equation}
Defining the conserved specific Killing energy
$\varepsilon\equiv\mathcal B/m_b$, Eq.~\eqref{eq:accBernoulli} gives
\begin{align}
 u(r)&=-\sqrt{\varepsilon^{2}-F(r)},\nonumber\\
 v(r)&=-\sqrt{1-\frac{F(r)}{\varepsilon^{2}}},\nonumber\\
 \rho(r)&=\frac{m_b|\mathcal C_N|}
 {r^{2}\sqrt{\varepsilon^{2}-F(r)}}.
\label{eq:accDustGeneral}
\end{align}
The negative roots describe inward motion.  Dust released from rest at
infinity has $u(\infty)=0$ and therefore $\varepsilon^{2}=c$, rather than
$\varepsilon^{2}=1$.  Its profiles reduce to
\begin{align}
 u(r)&=-\sqrt{c-F(r)}
 =-\sqrt{\frac{2M}{r}-\frac{\Qcal^{2}}{r^{2}}},
\label{eq:accDustU}\\
 v(r)&=-\sqrt{1-\frac{F(r)}{c}},
\label{eq:accDustV}\\
 \rho(r)&=\frac{m_b|\mathcal C_N|}
 {r^{2}\sqrt{c-F(r)}}.
\label{eq:accDustRho}
\end{align}
At the outer horizon, approached from $r>r_+$, these expressions have the
regular limits
\begin{align}
 u(r_+)&=-\sqrt c,
 &v(r)&\longrightarrow-1,\nonumber\\
 \rho(r_+)&=\frac{m_b|\mathcal C_N|}{r_+^{2}\sqrt c}.
\label{eq:accDustHorizon}
\end{align}
The static observer used to define $v$ does not exist on the horizon itself;
the second relation in Eq.~\eqref{eq:accDustHorizon} is an exterior limit.  At
large radius,
\begin{equation}
 |u|\simeq\sqrt{\frac{2M}{r}},
 \qquad
 |v|\simeq\sqrt{\frac{2M}{cr}},
 \qquad
 \rho\simeq\frac{m_b|\mathcal C_N|}{\sqrt{2M}\,r^{3/2}}.
\label{eq:accDustAsymptotic}
\end{equation}

This pressureless solution is a ballistic benchmark rather than a
finite-density Bondi--Michel boundary-value solution.  Indeed,
$n\propto r^{-3/2}$ implies $n_{\infty}=0$, the conserved flux remains an
independently prescribed normalization, and the vanishing sound speed means
that there is no finite sonic point.

For the illustrative plots we fix the integration constants $M=1$ and
$Q=0.6$, set $m_b=1$ and $\mathcal C_N=-10^{-3}$, and impose release from
rest at infinity, $\varepsilon=\sqrt c$.  We compare the
Reissner--Nordstr\"om reference $(l_1,l_2)=(0,0)$ with the two representative
Lorentz-violating choices $(0,0.05)$ and $(0.04,0.04)$.  The corresponding
triples $(c,\Qcal^{2},r_+)$ are
\begin{align}
 (0,0):&\quad (1,\,0.36,\,1.800000),\nonumber\\
 (0,0.05):&\quad (1.025641,\,0.378698,\,1.737492),\nonumber\\
 (0.04,0.04):&\quad (1.019608,\,0.391127,\,1.741231).
\label{eq:accBenchmarks}
\end{align}
These values hold at fixed metric integration constants $M$ and $Q$; they
should not be interpreted as a comparison at fixed independently measured
physical mass.  Plotting against $r/r_+$ separates the change in the horizon
location from the change in the exterior profile.

\begin{figure}[t]
 \includegraphics[width=\columnwidth]{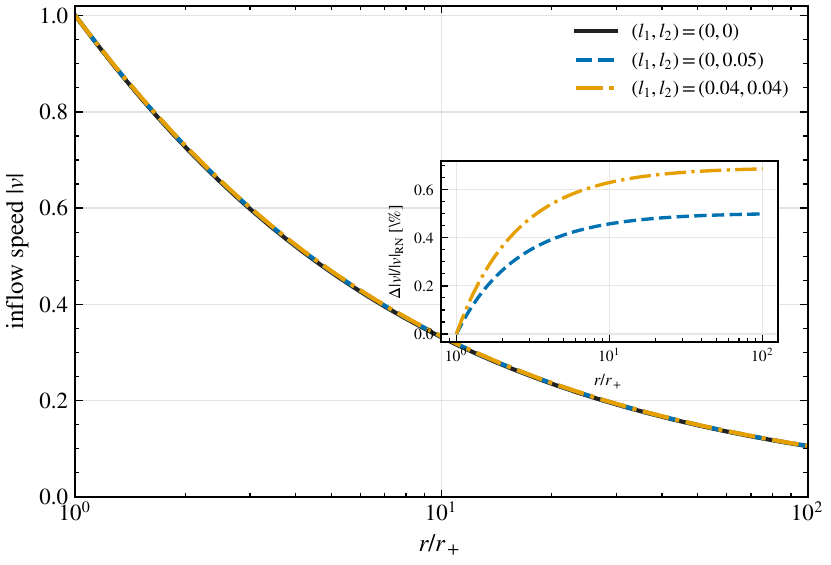}
 \caption{Magnitude of the local radial three-velocity $|v(r)|$ for steady
 pressureless infall released from rest at infinity, calculated from
 Eq.~\eqref{eq:accDustV}.  We use $M=1$, $Q=0.6$ and compare
 $(l_1,l_2)=(0,0)$, $(0,0.05)$, and $(0.04,0.04)$.  The horizontal variable
 is $r/r_+$.  The ingoing branch has $v<0$; each curve approaches $|v|=1$
 as $r\to r_+^{+}$ and approaches zero at large radius.  The inset shows the
 fractional shift relative to the Reissner--Nordstr\"om reference.}
 \label{fig:velocity}
\end{figure}

\begin{figure}[t]
 \includegraphics[width=\columnwidth]{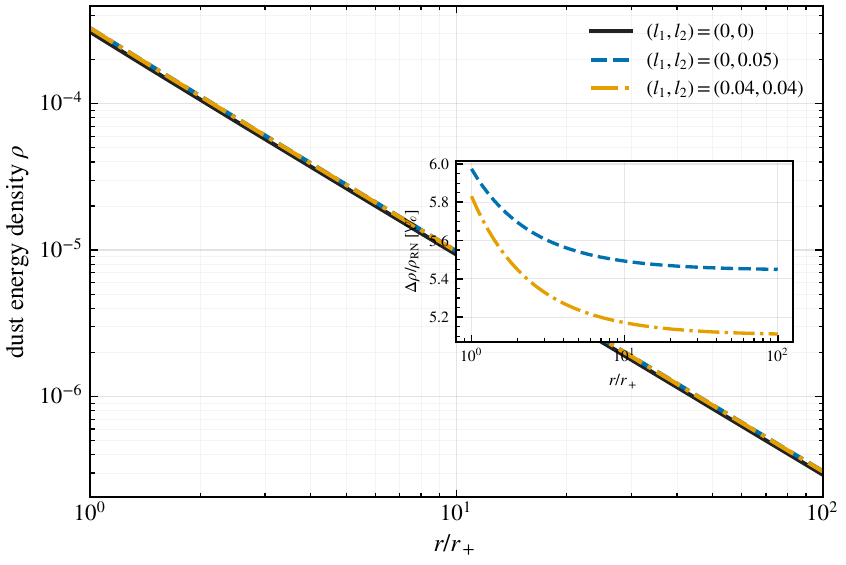}
 \caption{Rest-frame energy density of the same pressureless flow,
 $\rho=m_b n$, calculated from Eq.~\eqref{eq:accDustRho} for $m_b=1$ and
 $\mathcal C_N=-10^{-3}$.  The density remains finite at the outer horizon
 and falls as $r^{-3/2}$ at large radius.  Because the coupling-induced
 differences are modest, a fractional-residual inset relative to the
 Reissner--Nordstr\"om curve is useful.}
 \label{fig:density}
\end{figure}

For a regular Michel accretion branch with finite $n$ and $h$ at a
nonextremal horizon, $u$ remains finite and nonzero while the exterior static
observer measures $v\to-1$.  Hamiltonian level sets can also contain branches
with $v\to0$ at a horizon, as discussed in
~\cite{AhmedEtAl2016Accretion,AzregAinou2017Accretion}; by
Eq.~\eqref{eq:accDensityRV}, however, such branches generally require a
divergent number density and pressure.  They should therefore not be treated
as regular test-fluid accretion profiles without a separate analysis of
backreaction and fluid microphysics.  Finally, the present model neglects
angular momentum, viscosity, magnetic fields, radiative transfer, shocks,
and the self-gravity of the accreting gas.  It should be viewed as a
spherical test-flow diagnostic of the background geometry, not as
a complete accretion-disk model.

\section{Physical interpretation and limitations}
\label{sec:discussion}

\subsection{What the local angle measures}

The integration constant $M$ should not be identified without qualification with an asymptotically flat ADM mass.  Local ray dynamics measures $\mu=M/c^{3/2}$, and the charge correction measures $\Qbar^{2}=\Qcal^{2}/c^{2}$.  Consequently, a lensing analysis that imports an independently measured mass must translate that measurement into the same asymptotic normalization.  Otherwise a change attributed to $l_{2}$ may simply be a mismatch between mass conventions.

The leading local angle is achromatic in vacuum and proportional to $\mu/b$.  At linear order, $l_{2}$ is therefore almost perfectly degenerate with the mass.  The degeneracy can be weakened only by bringing in another scale: the $b^{-2}$ terms, the conical geometry, a photon-sphere observable, or a frequency-dependent plasma signal.  The $l_{1}$ direction is even more challenging because it enters the local weak angle only through $Q^{2}/b^{2}$ at this order.  For an effectively neutral astrophysical black hole, weak lensing alone does not probe $l_{1}$ linearly.

\subsection{Global cone versus local force}

The limit $\alphatot\rightarrow\pi(c^{-1/2}-1)$ as $b\rightarrow\infty$ is sometimes described loosely as a persistent deflection.  The more precise statement is that the original azimuth has a different global identification from the locally unwrapped coordinate.  A local observer comparing neighboring directions measures the decaying angle $\alphaloc$; a source-observer construction around the full object can also be sensitive to the global deficit or excess angle.

This distinction prevents a common overinterpretation.  Adding the constant term in Eq.~\eqref{eq:alphavaccoordinate} to a flat thin-lens equation would create a spurious universal image displacement.  The correct treatment must derive the lens map on the conical background or use a finite-distance definition in which the source, receiver, and optical triangle are all specified.  The present calculation supplies the local curvature input and the exact angle conversion needed for that next step, but it does not claim a global image-position constraint.

\subsection{Chromatic information and plasma degeneracy}

A homogeneous plasma increases the deflection at lower frequency, but Eq.~\eqref{eq:enhancementRatios} shows that the increase is sector dependent.  This is potentially useful because Lorentz-violating changes in $\mu$ and $\Qbar$ are achromatic, whereas the plasma weights vary with $\omega_{\infty}$.  In a controlled medium, measurements at several frequencies and impact parameters could therefore separate an overall mass renormalization from dispersive propagation.

The same frequency dependence also creates a degeneracy if the plasma model is not known.  An astrophysical plasma generally has $\omega_{p}=\omega_{p}(R)$, and its gradient produces an additional refractive deflection that is absent from Eq.~\eqref{eq:plasmaalpha} \cite{BisnovatyiKoganTsupko2010}.  The homogeneous result should therefore be regarded as an analytic benchmark and a code-validation limit, not as a complete model of an accretion environment.

\subsection{Relation to one-coupling calculations}

Setting $l_{1}=0$ and $l_{2}=2\ell$ recovers the charged one-parameter geometry used in several previous lensing studies.  The explicit reduction is given in Appendix~\ref{app:oneparameter}.  The two-coupling analysis clarifies which conclusions of that truncation are structural and which result from the parameter restriction.  In the one-coupling limit, the local charge parameter becomes independent of $\ell$, whereas in the full theory $l_{1}$ changes $\Qbar^{2}$ already at linear order.  This is why the two-dimensional response map contains a charge-sensitive direction absent from the reduced model.

The complete 2PM Gauss-Bonnet result also clarifies a methodological issue.  A straight-ray domain is sufficient for the leading weak angle and the leading charge correction, but not for the full mass-squared coefficient.  Agreement with the geodesic expansion is obtained only after the first-order trajectory is used as the integration boundary.  This point is independent of the Kalb-Ramond interpretation and applies more broadly to 2PM optical-curvature calculations.

Our results apply to the static, spherically symmetric, $\Lambda=0$ branch,
a fixed radial Kalb-Ramond vacuum, test-ray propagation, and either vacuum or
a homogeneous cold plasma.  We include the symmetric finite-distance weak-angle
correction derived above, but we do not include rotation, a complete finite-distance
lens map, perturbations of the Lorentz-violating field, nonuniform plasma profiles,
or backreaction of the medium.

The source theory admits branches with a cosmological constant, but a source/receiver-at-infinity formula is not appropriate in the presence of cosmological horizons or non-Euclidean asymptotics.  Those cases require a finite-distance optical-triangle construction \cite{IshiharaEtAl2016}.  A phenomenological extension should therefore proceed in the following order: construct the finite-distance conical lens equation, specify a mass and charge calibration, introduce a radial plasma profile, and only then compare with observational angular scales.

\section{Conclusions}
\label{sec:conclusions}

We have investigated weak gravitational deflection and steady spherical
test-fluid accretion in the static charged black-hole geometry generated by a
Lorentz-violating Kalb--Ramond background with two retained nonminimal
curvature couplings.  The optical results are conditional on treating the
observed radiation as a minimally coupled probe of the effective metric, and
the accretion analysis assumes a neutral, non-self-gravitating, stationary,
spherically symmetric, inviscid, and adiabatic fluid.  Within these
assumptions, the two calculations provide complementary diagnostics of the
same background: lensing probes its null optical geometry, whereas accretion
probes the timelike potential, the sphere area, and the normalization of the
conserved fluxes.

The first central result is the separation of local dynamics from global
asymptotic structure.  The metric does not approach the standard Minkowski
normalization in its original coordinates, so neither the coordinate time nor
the coordinate azimuth can be assigned their usual asymptotically flat
interpretation without qualification.  Canonical normalization converts the
local equatorial geometry into a Reissner--Nordstr\"om-like form with an
effective mass and charge, but it does not remove the nonstandard azimuthal
periodicity.  The background-subtracted local bending therefore decreases
with impact parameter, whereas the angle expressed in the original azimuth
retains a constant global contribution.  This constant is an angular
identification effect rather than a local gravitational force and cannot, by
itself, be interpreted as a universal image displacement.  Observable image
positions must instead be derived from a finite-distance lens construction
formulated directly on the conical background.

The second central result concerns the post-Minkowskian calculation.  In the
Gauss--Bonnet approach, both the conical large-circle boundary term and the
first-order displacement of the ray must be retained.  A straight-ray domain
reproduces the leading mass contribution and the leading charge correction,
but it misses part of the second-order mass term.  Incorporating the displaced
boundary restores the complete second-order coefficient and gives exact
agreement with the independent null-geodesic expansion.  The exact
turning-point integral and the third-order benchmark provide further checks,
demonstrating that the perturbative approximation improves systematically as
the scattering orbit moves away from the capture region.  This agreement also
confirms that the optical-curvature and orbit methods use the same
canonically normalized impact parameter and the same definition of the local
deflection angle.

The homogeneous-plasma extension introduces a genuinely chromatic response.
After the conserved frequency is normalized at infinity, the leading mass
term, the second-order mass term, and the charge term acquire different
frequency weights.  Plasma propagation therefore changes the structure of
the weak-field series rather than multiplying the vacuum angle by a universal
enhancement factor.  Exact Hamiltonian quadrature reproduces the analytic
coefficients and quantifies the truncation error.  The rapid increase of the
deflection near the propagation cutoff must nevertheless be considered
together with the strengthened weak-field requirements and the
energy-dependent unstable-orbit capture boundary.  The homogeneous model
should consequently be viewed as an analytic benchmark.  A realistic
accretion environment has a spatially varying plasma frequency and produces
an additional refractive-gradient contribution.

The hierarchy between the two Lorentz-violating couplings becomes transparent
only after the normalization convention is specified.  At fixed metric
integration constants, the second coupling changes the asymptotic cone and
the leading normalized mass sector, whereas the first coupling initially
enters through the effective charge and is therefore subleading in weak
lensing.  This explains why the numerical response is generally dominated by
the second coupling.  The hierarchy is nevertheless parametrization
dependent.  If the normalized conserved mass rather than the metric
integration constant is held fixed, the apparent linear dependence can be
redistributed between the two coupling directions.  No observational bound
should therefore be quoted until the integration constants, conserved mass,
electric charge, and asymptotic time normalization have been incorporated
into a single calibration scheme.

The accretion calculation supplies an independent consistency test of the
same normalization.  Particle conservation must be formulated in terms of
the baryon density rather than the total energy density.  Together with
Killing-energy conservation, it yields the conserved baryon flux, the
relativistic Bernoulli integral, and the Hamiltonian description of the flow.
For an isentropic barotropic fluid, regularity of the wind equation determines
the sonic-point conditions and identifies the smooth transonic branches.  In
canonically normalized coordinates, the local critical-point structure is
again Reissner--Nordstr\"om-like when the effective mass and charge are held
fixed.  The conical factor nevertheless remains in the physical area of the
symmetry spheres and therefore enters the integrated accretion rate.  The
coupling dependence of a coordinate sonic radius must consequently be
distinguished from the coupling dependence of a physically normalized flux.

The pressureless solution was used only as a controlled ballistic benchmark,
not as a complete Bondi--Michel gas model.  For dust released from rest at
infinity, the radial four-velocity and rest-frame density remain finite at the
outer horizon.  The speed measured by exterior static observers approaches
the speed of light in the horizon limit, while the density decreases with the
expected power-law behavior far from the black hole.  The changes generated
by the Lorentz-violating couplings are modest for the benchmark parameters and
are evaluated at fixed metric mass and charge.  These profiles should not be
interpreted as predictions for a luminous accretion disk.  A physical
transonic model additionally requires an equation of state, asymptotic
density and temperature, sonic regularity, and ultimately radiative and
magnetic transport.

Taken together, the results separate the observable content of the geometry
into global conical, local mass, local charge, dispersive plasma, and fluid
critical-point sectors.  Because these sectors respond differently to impact
parameter, frequency, and asymptotic normalization, they may eventually help
break parameter degeneracies.  Such an interpretation is possible only after
the mass definition, probe field, plasma distribution, and fluid boundary
conditions have been fixed consistently.  Natural extensions include a
finite-distance lens equation constructed on the conical background,
polarization-dependent geometric optics for the nonminimally coupled
electromagnetic sector, ray tracing through an inhomogeneous plasma, fully
transonic accretion including backreaction, and rotating Kalb--Ramond
geometries.  These developments are necessary before the analytic signatures
identified here can be converted into robust observational constraints on
the two Lorentz-violating couplings.

\begin{acknowledgments}
A. \"O and R. P. would like to acknowledge the networking support of the COST Action CA21106 - COSMIC WISPers in the Dark Universe: Theory, astrophysics and experiments (CosmicWISPers), the COST Action CA22113 - Fundamental challenges in theoretical physics (THEORY-CHALLENGES), the COST Action CA21136 - Addressing observational tensions in cosmology with systematics and fundamental physics (CosmoVerse), the COST Action CA23130 - Bridging high and low energies in search of quantum gravity (BridgeQG), and the COST Action CA23115 - Relativistic Quantum Information (RQI) funded by COST (European Cooperation in Science and Technology).
\end{acknowledgments}

\section*{Data Availability Statement}
No external observational data were used.  

\appendix

\section{Second-order Gauss-Bonnet integrations}
\label{app:gbt}

The radial integral of the vacuum curvature density outside the ray is
\begin{equation}
 \int_{R_{\gamma}}^{\infty}K\dd S
 =-2\mu U_{\gamma}
 +\frac{3}{2}(\Qbar^{2}-\mu^{2})U_{\gamma}^{2}
 +\Order(\epsilon^{3}).
\label{eq:radialintegralvac}
\end{equation}
The explicitly second-order term needs only the zeroth-order orbit, whereas the leading mass term must be evaluated on the first-order shifted boundary.

Writing $U_{\gamma}=U_{0}+\mu U_{1}$, the required angular integrals are
\begin{equation}
 \int_{0}^{\pi}U_{0}\dd\varphi=\frac{2}{b},
 \quad
 \int_{0}^{\pi}U_{1}\dd\varphi=\frac{3\pi}{2b^{2}},
 \quad
 \int_{0}^{\pi}U_{0}^{2}\dd\varphi=\frac{\pi}{2b^{2}}.
\label{eq:vacangularintegrals}
\end{equation}
The middle integral is the boundary-displacement contribution that completes the $15\pi/4$ coefficient in Eq.~\eqref{eq:alphavaclocal}.

For the plasma calculation, define the frequency-dependent combinations
\begin{equation}
\begin{aligned}
 P_{v}&=1+v^{-2},\\
 C_{v}&=\Qbar^{2}(1+2v^{-2})
 +\mu^{2}(-1-6v^{-2}+4v^{-4}).
\end{aligned}
\label{eq:PvCv}
\end{equation}
The Gauss-Bonnet integral then takes the compact form
\begin{equation}
 \alpha_{\mathrm{loc}}^{p}
 =\int_{0}^{\pi}\left[\mu P_{v}U_{\gamma}^{p}
 -\frac{C_{v}}{2}U_{0}^{2}\right]\dd\varphi.
\label{eq:plasmaGBTint}
\end{equation}
The remaining trajectory integral is
\begin{equation}
 \int_{0}^{\pi}\frac{v^{-2}+\cos^{2}\varphi}{b^{2}}
 \dd\varphi
 =\frac{\pi}{b^{2}}\left(v^{-2}+\frac{1}{2}\right).
\label{eq:plasmaTrajectoryIntegral}
\end{equation}
Substitution of Eqs.~\eqref{eq:PvCv} and \eqref{eq:plasmaTrajectoryIntegral} reduces Eq.~\eqref{eq:plasmaGBTint} to the plasma angle in Eq.~\eqref{eq:plasmaalpha}.

\section{One-parameter limit}
\label{app:oneparameter}

The charged one-coupling geometry is obtained from the two-coupling branch by setting
\begin{equation}
 l_{1}=0,
 \qquad
 l_{2}=2\ell.
\label{eq:onelimit}
\end{equation}
The normalized parameters then reduce to
\begin{align}
 c&=\frac{1}{1-\ell},
 &
 \Qcal^{2}&=\frac{Q^{2}}{(1-\ell)^{2}},
\nonumber\\
 \mu&=M(1-\ell)^{3/2},
 &
 \Qbar^{2}&=Q^{2}.
\label{eq:onemap}
\end{align}
The charge sector is therefore unchanged in the local normalized geometry, while the mass and cone retain the Lorentz-violating dependence.

The background-subtracted 2PM angle becomes
\begin{equation}
 \alphaloc=\frac{4M(1-\ell)^{3/2}}{b}
 +\frac{15\pi M^{2}(1-\ell)^{3}}{4b^{2}}
 -\frac{3\pi Q^{2}}{4b^{2}}.
\label{eq:onealpha}
\end{equation}
Formulas expressed in the original coordinate angle differ by the conversion in Eq.~\eqref{eq:anglerelation}, which adds $\pi[\sqrt{1-\ell}-1]$ and multiplies the local angle by $\sqrt{1-\ell}$.

\bibliography{ref}

@article{KalbRamond1974,
 author = "Kalb, Michael and Ramond, Pierre",
    title = "{Classical direct interstring action}",
    doi = "10.1103/PhysRevD.9.2273",
    journal = "Phys. Rev. D",
    volume = "9",
    pages = "2273--2284",
    year = "1974"
}

@article{KosteleckySamuel1989,
 author = "Kostelecky, V. Alan and Samuel, Stuart",
    title = "{Spontaneous Breaking of Lorentz Symmetry in String Theory}",
    reportNumber = "IUHET-139, CCNY-HEP-88/4",
    doi = "10.1103/PhysRevD.39.683",
    journal = "Phys. Rev. D",
    volume = "39",
    pages = "683",
    year = "1989"
}

@article{AltschulBaileyKostelecky2010,
 author = "Altschul, Brett and Bailey, Quentin G. and Kostelecky, V. Alan",
    title = "{Lorentz violation with an antisymmetric tensor}",
    eprint = "0912.4852",
    archivePrefix = "arXiv",
    primaryClass = "gr-qc",
    reportNumber = "IUHET-537",
    doi = "10.1103/PhysRevD.81.065028",
    journal = "Phys. Rev. D",
    volume = "81",
    pages = "065028",
    year = "2010"
}

@article{GibbonsWerner2008,
    author = "Gibbons, G. W. and Werner, M. C.",
    title = "{Applications of the Gauss-Bonnet theorem to gravitational lensing}",
    eprint = "0807.0854",
    archivePrefix = "arXiv",
    primaryClass = "gr-qc",
    doi = "10.1088/0264-9381/25/23/235009",
    journal = "Class. Quant. Grav.",
    volume = "25",
    pages = "235009",
    year = "2008"
}

@article{Werner2012,
   author = "Werner, M. C.",
    title = "{Gravitational lensing in the Kerr-Randers optical geometry}",
    eprint = "1205.3876",
    archivePrefix = "arXiv",
    primaryClass = "gr-qc",
    doi = "10.1007/s10714-012-1458-9",
    journal = "Gen. Rel. Grav.",
    volume = "44",
    pages = "3047--3057",
    year = "2012"
}

@article{IshiharaEtAl2016,
    author = "Ishihara, Asahi and Suzuki, Yusuke and Ono, Toshiaki and Kitamura, Takao and Asada, Hideki",
    title = "{Gravitational bending angle of light for finite distance and the Gauss-Bonnet theorem}",
    eprint = "1604.08308",
    archivePrefix = "arXiv",
    primaryClass = "gr-qc",
    doi = "10.1103/PhysRevD.94.084015",
    journal = "Phys. Rev. D",
    volume = "94",
    number = "8",
    pages = "084015",
    year = "2016"
}

@article{CrisnejoGallo2018,
  author = "Crisnejo, Gabriel and Gallo, Emanuel",
    title = "{Weak lensing in a plasma medium and gravitational deflection of massive particles using the Gauss-Bonnet theorem. A unified treatment}",
    eprint = "1804.05473",
    archivePrefix = "arXiv",
    primaryClass = "gr-qc",
    doi = "10.1103/PhysRevD.97.124016",
    journal = "Phys. Rev. D",
    volume = "97",
    number = "12",
    pages = "124016",
    year = "2018"
}

@article{PerlickTsupkoBisnovatyiKogan2015,
    author = "Perlick, Volker and Tsupko, Oleg Yu. and Bisnovatyi-Kogan, Gennady S.",
    title = "{Influence of a plasma on the shadow of a spherically symmetric black hole}",
    eprint = "1507.04217",
    archivePrefix = "arXiv",
    primaryClass = "gr-qc",
    doi = "10.1103/PhysRevD.92.104031",
    journal = "Phys. Rev. D",
    volume = "92",
    number = "10",
    pages = "104031",
    year = "2015"
}

@article{BisnovatyiKoganTsupko2010,
    author = "Bisnovatyi-Kogan, G. S. and Tsupko, O. Yu.",
    title = "{Gravitational lensing in a non-uniform plasma}",
    eprint = "1006.2321",
    archivePrefix = "arXiv",
    primaryClass = "astro-ph.CO",
    doi = "10.1111/j.1365-2966.2010.16290.x",
    journal = "Mon. Not. Roy. Astron. Soc.",
    volume = "404",
    pages = "1790--1800",
    year = "2010"
}

@article{FuZhaoLiu2021,
    author = "Fu, Qi-Ming and Zhao, Li and Liu, Yu-Xiao",
    title = "{Weak deflection angle by electrically and magnetically charged black holes from nonlinear electrodynamics}",
    eprint = "2101.08409",
    archivePrefix = "arXiv",
    primaryClass = "gr-qc",
    doi = "10.1103/PhysRevD.104.024033",
    journal = "Phys. Rev. D",
    volume = "104",
    number = "2",
    pages = "024033",
    year = "2021"
}

@article{JuniorEtAl2024,
   author = "Junior, Ednaldo L. B. and Junior, Jos{\'e} Tarciso S. S. and Lobo, Francisco S. N. and Rodrigues, Manuel E. and Rubiera-Garcia, Diego and da Silva, Lu{\'\i}s F. Dias and Vieira, Henrique A.",
    title = "{Gravitational lensing of a Schwarzschild-like black hole in Kalb-Ramond gravity}",
    eprint = "2405.03284",
    archivePrefix = "arXiv",
    primaryClass = "gr-qc",
    doi = "10.1103/PhysRevD.110.024077",
    journal = "Phys. Rev. D",
    volume = "110",
    number = "2",
    pages = "024077",
    year = "2024"
}

@article{PantigOvgunRincon2025,
   author = {Pantig, Reggie C. and {\"O}vg{\"u}n, Ali and Rinc{\'o}n, {\'A}ngel},
    title = "{Charged black holes in KR gravity: Weak deflection angle, shadow cast, quasinormal modes and neutrino annihilation}",
    eprint = "2505.17947",
    archivePrefix = "arXiv",
    primaryClass = "gr-qc",
    doi = "10.1016/j.dark.2025.102029",
    journal = "Phys. Dark Univ.",
    volume = "49",
    pages = "102029",
    year = "2025"
}

@article{TangLan2025,
  author = "Tan, K. and Lan, X. G.",
    title = "{Charged black holes in the Kalb-Ramond background with Lorentz violation: null geodesics and optical appearance of a thin accretion disk*}",
    eprint = "2503.16800",
    archivePrefix = "arXiv",
    primaryClass = "gr-qc",
    doi = "10.1088/1674-1137/adc3fc",
    journal = "Chin. Phys. C",
    volume = "49",
    number = "7",
    pages = "075104",
    year = "2025"
}

@article{AraujoFilhoEtAl2024,
  author = "Ara{\'u}jo Filho, A. A. and Heidari, N. and Reis, J. A. A. S. and Hassanabadi, H.",
    title = "{The impact of an antisymmetric tensor on charged black holes: evaporation process, geodesics, deflection angle, scattering effects and quasinormal modes}",
    eprint = "2404.10721",
    archivePrefix = "arXiv",
    primaryClass = "gr-qc",
    doi = "10.1088/1361-6382/adbb4f",
    journal = "Class. Quant. Grav.",
    volume = "42",
    number = "6",
    pages = "065026",
    year = "2025"
}

@article{EiroaRomeroTorres2002,
    author = "Eiroa, Ernesto F. and Romero, Gustavo E. and Torres, Diego F.",
    title = "{Reissner-Nordstrom black hole lensing}",
    eprint = "gr-qc/0203049",
    archivePrefix = "arXiv",
    doi = "10.1103/PhysRevD.66.024010",
    journal = "Phys. Rev. D",
    volume = "66",
    pages = "024010",
    year = "2002"
}

@article{KeetonPetters2005,
  author = "Keeton, Charles R. and Petters, A. O.",
    title = "{Formalism for testing theories of gravity using lensing by compact objects. I. Static, spherically symmetric case}",
    eprint = "gr-qc/0511019",
    archivePrefix = "arXiv",
    doi = "10.1103/PhysRevD.72.104006",
    journal = "Phys. Rev. D",
    volume = "72",
    pages = "104006",
    year = "2005"
}

@article{Bozza2002,
   author = "Bozza, V.",
    title = "{Gravitational lensing in the strong field limit}",
    eprint = "gr-qc/0208075",
    archivePrefix = "arXiv",
    doi = "10.1103/PhysRevD.66.103001",
    journal = "Phys. Rev. D",
    volume = "66",
    pages = "103001",
    year = "2002"
}

@article{Kostelecky:1989jw,
    author = "Kostelecky, V. Alan and Samuel, Stuart",
    title = "{Gravitational Phenomenology in Higher Dimensional Theories and Strings}",
    reportNumber = "IUHET-157",
    doi = "10.1103/PhysRevD.40.1886",
    journal = "Phys. Rev. D",
    volume = "40",
    pages = "1886--1903",
    year = "1989"
}

@article{Carroll:2001ws,
   author = "Carroll, Sean M. and Harvey, Jeffrey A. and Kostelecky, V. Alan and Lane, Charles D. and Okamoto, Takemi",
    title = "{Noncommutative field theory and Lorentz violation}",
    eprint = "hep-th/0105082",
    archivePrefix = "arXiv",
    reportNumber = "EFI-01-12, IUHET-433",
    doi = "10.1103/PhysRevLett.87.141601",
    journal = "Phys. Rev. Lett.",
    volume = "87",
    pages = "141601",
    year = "2001"
}

@article{Kostelecky:2003fs,
  author = "Kostelecky, V. Alan",
    title = "{Gravity, Lorentz violation, and the standard model}",
    eprint = "hep-th/0312310",
    archivePrefix = "arXiv",
    reportNumber = "IUHET-461",
    doi = "10.1103/PhysRevD.69.105009",
    journal = "Phys. Rev. D",
    volume = "69",
    pages = "105009",
    year = "2004"
}

@article{Higashijima:2001sq,
   author = "Higashijima, Kiyoshi and Yokoi, Naoto",
    title = "{Spontaneous Lorentz symmetry breaking by antisymmetric tensor field}",
    eprint = "hep-th/0101222",
    archivePrefix = "arXiv",
    reportNumber = "OU-HET-377",
    doi = "10.1103/PhysRevD.64.025004",
    journal = "Phys. Rev. D",
    volume = "64",
    pages = "025004",
    year = "2001"
}

@article{Maluf:2018jwc,
   author = "Maluf, R. V. and Ara{\'u}jo Filho, A. A. and Cruz, W. T. and Almeida, C. A. S.",
    title = "{Antisymmetric tensor propagator with spontaneous Lorentz violation}",
    eprint = "1810.04003",
    archivePrefix = "arXiv",
    primaryClass = "hep-th",
    doi = "10.1209/0295-5075/124/61001",
    journal = "EPL",
    volume = "124",
    number = "6",
    pages = "61001",
    year = "2018"
}

@article{Majumdar:1999jd,
  author = "Majumdar, Parthasarathi and SenGupta, Soumitra",
    title = "{Parity violating gravitational coupling of electromagnetic fields}",
    eprint = "gr-qc/9906027",
    archivePrefix = "arXiv",
    doi = "10.1088/0264-9381/16/12/102",
    journal = "Class. Quant. Grav.",
    volume = "16",
    pages = "L89--L94",
    year = "1999"
}

@article{Lessa:2019bgi,
   author = "Lessa, L. A. and Silva, J. E. G. and Maluf, R. V. and Almeida, C. A. S.",
    title = "{Modified black hole solution with a background Kalb{\textendash}Ramond field}",
    eprint = "1911.10296",
    archivePrefix = "arXiv",
    primaryClass = "gr-qc",
    doi = "10.1140/epjc/s10052-020-7902-1",
    journal = "Eur. Phys. J. C",
    volume = "80",
    number = "4",
    pages = "335",
    year = "2020"
}

@article{Lessa:2020imi,
    author = "Lessa, L. A. and Oliveira, R. and Silva, J. E. G. and Almeida, C. A. S.",
    title = "{Traversable wormhole solution with a background Kalb{\textendash}Ramond field}",
    eprint = "2010.05298",
    archivePrefix = "arXiv",
    primaryClass = "gr-qc",
    doi = "10.1016/j.aop.2021.168604",
    journal = "Annals Phys.",
    volume = "433",
    pages = "168604",
    year = "2021"
}

@article{Maluf:2021eyu,
   author = "Maluf, R. V. and Muniz, C. R.",
    title = "{Exact solution for a traversable wormhole in a curvature-coupled antisymmetric background field}",
    eprint = "2110.12202",
    archivePrefix = "arXiv",
    primaryClass = "gr-qc",
    doi = "10.1140/epjc/s10052-022-10409-7",
    journal = "Eur. Phys. J. C",
    volume = "82",
    number = "5",
    pages = "445",
    year = "2022"
}

@article{Atamurotov:2022slw,
    author = "Atamurotov, Farruh and Ortiqboev, Dilmurod and Abdujabbarov, Ahmadjon and Mustafa, G.",
    title = "{Particle dynamics and gravitational weak lensing around black hole in the Kalb-Ramond gravity}",
    doi = "10.1140/epjc/s10052-022-10619-z",
    journal = "Eur. Phys. J. C",
    volume = "82",
    number = "8",
    pages = "659",
    year = "2022"
}

@article{Yang:2023wtu,
    author = "Yang, Ke and Chen, Yue-Zhe and Duan, Zheng-Qiao and Zhao, Ju-Ying",
    title = "{Static and spherically symmetric black holes in gravity with a background Kalb-Ramond field}",
    eprint = "2308.06613",
    archivePrefix = "arXiv",
    primaryClass = "gr-qc",
    doi = "10.1103/PhysRevD.108.124004",
    journal = "Phys. Rev. D",
    volume = "108",
    number = "12",
    pages = "124004",
    year = "2023"
}

@article{Duan:2023qsg,
     author = "Duan, Zheng-Qiao and Zhao, Ju-Ying and Yang, Ke",
    title = "{Electrically charged black holes in gravity with a background Kalb{\textendash}Ramond field}",
    eprint = "2310.13555",
    archivePrefix = "arXiv",
    primaryClass = "gr-qc",
    doi = "10.1140/epjc/s10052-024-13188-5",
    journal = "Eur. Phys. J. C",
    volume = "84",
    number = "8",
    pages = "798",
    year = "2024"
}

@article{Liu:2024gxr,
    author = "Liu, Wentao and Wu, Di and Wang, Jieci",
    title = "{Static neutral black holes in Kalb-Ramond gravity}",
    eprint = "2406.13461",
    archivePrefix = "arXiv",
    primaryClass = "hep-th",
    doi = "10.1088/1475-7516/2024/09/017",
    journal = "JCAP",
    volume = "09",
    pages = "017",
    year = "2024"
}

@article{AraujoFilho:2023qea,
   author = "Filho, A. A. Ara{\'u}jo and Reis, J. A. A. S. and Hassanabadi, H.",
    title = "{Exploring antisymmetric tensor effects on black hole shadows and quasinormal frequencies}",
    eprint = "2309.15778",
    archivePrefix = "arXiv",
    primaryClass = "gr-qc",
    doi = "10.1088/1475-7516/2024/05/029",
    journal = "JCAP",
    volume = "05",
    pages = "029",
    year = "2024"
}

@article{Liu:2024yhu,
    author = "Liu, Wentao and Wu, Di and Wang, Jieci",
    title = "{Shadow of slowly rotating Kalb-Ramond black holes}",
    eprint = "2407.07416",
    archivePrefix = "arXiv",
    primaryClass = "gr-qc",
    doi = "10.1088/1475-7516/2025/05/017",
    journal = "JCAP",
    volume = "05",
    pages = "017",
    year = "2025"
}

@article{Liu:2025KR,
  author = "Liu, Jia-Zhou and Wu, Shan-Ping and Wei, Shao-Wen and Liu, Yu-Xiao",
    title = "{Exact black hole solutions in gravity with a background Kalb-Ramond field}",
    eprint = "2505.07404",
    archivePrefix = "arXiv",
    primaryClass = "gr-qc",
    doi = "10.1088/1475-7516/2025/11/056",
    journal = "JCAP",
    volume = "11",
    pages = "056",
    year = "2025"
}

@article{Xia:2025KR,
   author = "Xia, Zhong-Wu and Long, Sheng and Gong, Huajie and Pan, Qiyuan and Jing, Jiliang",
    title = "{Scalar perturbation around a rotating Kalb-Ramond BTZ black hole}",
    eprint = "2511.00784",
    archivePrefix = "arXiv",
    primaryClass = "gr-qc",
    doi = "10.1007/s11433-025-2921-y",
    journal = "Sci. China Phys. Mech. Astron.",
    volume = "69",
    number = "6",
    pages = "260411",
    year = "2026"
}

@article{Deng:2025KR,
  author = "Deng, Weike and Liu, Wentao and Xiao, Kui and Jing, Jiliang",
    title = "{Quasinormal modes of scalar, electromagnetic, and gravitational perturbations in slowly rotating Kalb{\textendash}Ramond black holes}",
    eprint = "2511.19553",
    archivePrefix = "arXiv",
    primaryClass = "gr-qc",
    doi = "10.1140/epjc/s10052-026-15470-0",
    journal = "Eur. Phys. J. C",
    volume = "86",
    number = "3",
    pages = "232",
    year = "2026"
}

@article{Lin:2026KR,
   author = "Lin, Yu-Xuan and Liu, Jia-Zhou and Liu, Yu-Xiao",
    title = "{Dyonic Black Holes in Lorentz-Violating Gravity with a Background Kalb--Ramond Field}",
    eprint = "2605.18371",
    archivePrefix = "arXiv",
    primaryClass = "gr-qc",
    month = "5",
    year = "2026"
}

@article{Yang:2026KR,
  author = "Yang, Jia-Hui and Guo, Xin-Yu and Liu, Jia-Zhou and Liu, Yu-Xiao",
    title = "{Charged Black Holes with a Lorentz--Violating Kalb--Ramond Background}",
    eprint = "2608.02196",
    archivePrefix = "arXiv",
    primaryClass = "gr-qc",
    month = "8",
    year = "2026"
}

@article{Battista:2026nsx,
    author = "Battista, Emmanuele and Capozziello, Salvatore and Chen, Che-Yu",
    title = "{Shadow signatures and energy accumulation in Lorentzian-Euclidean black holes}",
    eprint = "2601.10806",
    archivePrefix = "arXiv",
    primaryClass = "gr-qc",
    reportNumber = "RIKEN-iTHEMS-Report-26",
    doi = "10.1103/zf2w-7fqn",
    journal = "Phys. Rev. D",
    volume = "113",
    number = "10",
    pages = "104039",
    year = "2026"
}

@article{Vagnozzi:2022moj,
    author = "Vagnozzi, Sunny and others",
    title = "{Horizon-scale tests of gravity theories and fundamental physics from the Event Horizon Telescope image of Sagittarius A}",
    eprint = "2205.07787",
    archivePrefix = "arXiv",
    primaryClass = "gr-qc",
    reportNumber = "UCI-HEP-TR-2022-07",
    doi = "10.1088/1361-6382/acd97b",
    journal = "Class. Quant. Grav.",
    volume = "40",
    number = "16",
    pages = "165007",
    year = "2023"
}

@article{Kumar:2020hgm,
    author = "Kumar, Rahul and Ghosh, Sushant G. and Wang, Anzhong",
    title = "{Gravitational deflection of light and shadow cast by rotating Kalb-Ramond black holes}",
    eprint = "2001.00460",
    archivePrefix = "arXiv",
    primaryClass = "gr-qc",
    doi = "10.1103/PhysRevD.101.104001",
    journal = "Phys. Rev. D",
    volume = "101",
    number = "10",
    pages = "104001",
    year = "2020"
}

@article{Murodov:2026vmd,
    author = "Murodov, Sardor and Kholturayev, Olimjon and Ahmedov, Bahodir and Rahmatov, Bekzod and Egamberdiev, Islom and Ahmedov, Bobomurat",
    title = "{Massive neutral Dirac quasibound states in a Newman-Janis-generated rotating charged Kalb-Ramond black-hole geometry}",
    eprint = "2608.09313",
    archivePrefix = "arXiv",
    primaryClass = "gr-qc",
    month = "8",
    year = "2026"
}

@article{Battista:2026rtl,
    author = "Battista, Emmanuele and Campagnola, Roberto and Capozziello, Salvatore and Fiorillo, Giuseppe",
    title = "{Equivalence principle violation in metric-affine gravity and finite-temperature effects}",
    eprint = "2606.02329",
    archivePrefix = "arXiv",
    primaryClass = "gr-qc",
    doi = "10.1103/7qp5-9kzk",
    journal = "Phys. Rev. D",
    volume = "114",
    number = "2",
    pages = "024009",
    year = "2026"
}

@article{Nengroo:2026iju,
    author = "Nengroo, Towheed Ahmad and Islam, Shafqat Ul and Ghosh, Sushant G.",
    title = "{Probing Kalb-Ramond gravity with charged rotating black holes: constraints from EHT observations}",
    eprint = "2604.13494",
    archivePrefix = "arXiv",
    primaryClass = "gr-qc",
    month = "4",
    year = "2026"
}

@article{Shodikulov:2025xax,
    author = "Shodikulov, Bakhodir and Mirov, Mirjavokhir and Atamurotov, Farruh and Ghosh, Sushant G. and Abdujabbarov, Ahmadjon",
    title = "{Impact of Kalb{\textendash}Ramond fields and perfect fluid dark matter on black hole shadows and gravitational lensing}",
    doi = "10.1016/j.dark.2025.102096",
    journal = "Phys. Dark Univ.",
    volume = "50",
    pages = "102096",
    year = "2025"
}

@article{Alfaro2002,
   author = "Alfaro, Jorge and Morales-Tecotl, Hugo A. and Urrutia, Luis F.",
    title = "{Loop quantum gravity and light propagation}",
    eprint = "hep-th/0108061",
    archivePrefix = "arXiv",
    doi = "10.1103/PhysRevD.65.103509",
    journal = "Phys. Rev. D",
    volume = "65",
    pages = "103509",
    year = "2002"
}

@article{Liu2026BumblebeeExact,
     author = "Liu, Jia-Zhou and Wu, Shan-Ping and Wei, Shao-Wen and Liu, Yu-Xiao",
    title = "{Exact black hole solutions in bumblebee gravity with lightlike or spacelike VEVs}",
    eprint = "2510.16731",
    archivePrefix = "arXiv",
    primaryClass = "gr-qc",
    doi = "10.1007/s11433-026-2961-8",
    journal = "Sci. China Phys. Mech. Astron.",
    volume = "69",
    number = "7",
    pages = "270411",
    year = "2026"
}

@article{Liu2025ChargedBumblebee,
   author = "Liu, Jia-Zhou and Guo, Wen-Di and Wei, Shao-Wen and Liu, Yu-Xiao",
    title = "{Charged spherically symmetric and slowly rotating charged black hole solutions in bumblebee gravity}",
    eprint = "2407.08396",
    archivePrefix = "arXiv",
    primaryClass = "gr-qc",
    doi = "10.1140/epjc/s10052-025-13859-x",
    journal = "Eur. Phys. J. C",
    volume = "85",
    number = "2",
    pages = "145",
    year = "2025"
}

@article{An2024Bumblebee,
   author = "An, Yu-Sen",
    title = "{Notes on thermodynamics of Schwarzschild-like bumblebee black hole}",
    eprint = "2401.15430",
    archivePrefix = "arXiv",
    primaryClass = "gr-qc",
    doi = "10.1016/j.dark.2024.101520",
    journal = "Phys. Dark Univ.",
    volume = "45",
    pages = "101520",
    year = "2024"
}

@article{Liu2023BumblebeeQNM,
   author = "Liu, Wentao and Fang, Xiongjun and Jing, Jiliang and Wang, Jieci",
    title = "{QNMs of slowly rotating Einstein{\textendash}Bumblebee black hole}",
    eprint = "2211.03156",
    archivePrefix = "arXiv",
    primaryClass = "gr-qc",
    doi = "10.1140/epjc/s10052-023-11231-5",
    journal = "Eur. Phys. J. C",
    volume = "83",
    number = "1",
    pages = "83",
    year = "2023"
}

@article{Deng2025Bumblebee,
  author = "Deng, Weike and Liu, Wentao and Long, Fen and Xiao, Kui and Jing, Jiliang",
    title = "{Quasinormal modes of a massive scalar field in slowly rotating Einstein-Bumblebee black holes}",
    eprint = "2507.13978",
    archivePrefix = "arXiv",
    primaryClass = "gr-qc",
    doi = "10.1088/1475-7516/2025/11/028",
    journal = "JCAP",
    volume = "11",
    pages = "028",
    year = "2025"
}

@article{Li2025BumblebeeQNM,
    author = "Li, Bo-Rui and Liu, Jia-Zhou and Guo, Wen-Di and Liu, Yu-Xiao",
    title = "{Quasinormal modes of a charged spherically symmetric black hole in bumblebee gravity}",
    eprint = "2510.20503",
    archivePrefix = "arXiv",
    primaryClass = "gr-qc",
    month = "10",
    year = "2025"
}

@article{Singh2025,
   author = "Singh, Yenshembam Priyobarta and Devi, Irengbam Roshila and Singh, Telem Ibungochouba",
    title = "{Quasinormal modes of spherically symmetric black hole with cosmological constant and global monopole in bumblebee gravity}",
    eprint = "2504.09108",
    archivePrefix = "arXiv",
    primaryClass = "gr-qc",
    doi = "10.1016/j.nuclphysb.2025.117006",
    journal = "Nucl. Phys. B",
    volume = "1018",
    pages = "117006",
    year = "2025"
}

@article{Liu2024Isospectrality,
   author = "Liu, Wentao and Fang, Xiongjun and Jing, Jiliang and Wang, Jieci",
    title = "{Lorentz violation induces isospectrality breaking in Einstein-bumblebee gravity theory}",
    eprint = "2402.09686",
    archivePrefix = "arXiv",
    primaryClass = "gr-qc",
    doi = "10.1007/s11433-024-2405-y",
    journal = "Sci. China Phys. Mech. Astron.",
    volume = "67",
    number = "8",
    pages = "280413",
    year = "2024"
}

@article{Liu2026Decoupling,
 author = "Liu, Hui-Fa and Liu, Wentao and Liu, Yu-Xiao and Su, Qi and Zeng, Ding-fang",
    title = "{Gravitational-bumblebee perturbations: exact decoupling and isospectrality}",
    eprint = "2605.02820",
    archivePrefix = "arXiv",
    primaryClass = "gr-qc",
    doi = "10.1007/JHEP08(2026)027",
    journal = "JHEP",
    volume = "08",
    pages = "027",
    year = "2026"
}

@article{Lai2026,
  author = "Lai, Xiao-Bin and Dong, Yu-Qi and Fan, Yu-Zhi and Liu, Yu-Xiao",
    title = "{Stability analysis of cosmological perturbations in the bumblebee model: Parameter constraints and gravitational waves}",
    eprint = "2509.13958",
    archivePrefix = "arXiv",
    primaryClass = "gr-qc",
    doi = "10.1103/q6fk-3lkj",
    journal = "Phys. Rev. D",
    volume = "113",
    number = "4",
    pages = "044003",
    year = "2026"
}

@article{Liu2025BTZ,
    author = "Liu, Xiaofang and Liu, Wentao and Liu, Zhilong and Wang, Jieci",
    title = "{Harvesting correlations from BTZ black hole coupled to a Lorentz-violating vector field}",
    eprint = "2503.06404",
    archivePrefix = "arXiv",
    primaryClass = "gr-qc",
    doi = "10.1007/JHEP08(2025)094",
    journal = "JHEP",
    volume = "08",
    pages = "094",
    year = "2025"
}

@article{Tang2025,
 author = "Tang, Yu and Liu, Wentao and Wang, Jieci",
    title = "{Observational signature of Lorentz violation in acceleration radiation}",
    eprint = "2502.03043",
    archivePrefix = "arXiv",
    primaryClass = "gr-qc",
    doi = "10.1140/epjc/s10052-025-14797-4",
    journal = "Eur. Phys. J. C",
    volume = "85",
    number = "10",
    pages = "1108",
    year = "2025"
}

@article{GuChargedKRQNM,
   author = "Gu, Yun-Tao and Guo, Wen-Di and Liu, Yu-Xiao",
    title = "{Quasinormal modes of an electrically charged Kalb-Ramond black hole}",
    eprint = "2509.23732",
    archivePrefix = "arXiv",
    primaryClass = "gr-qc",
    month = "9",
    year = "2025"
}

@article{Bondi1952Accretion,
    author = "Bondi, H.",
    title = "{On spherically symmetrical accretion}",
    doi = "10.1093/mnras/112.2.195",
    journal = "Mon. Not. Roy. Astron. Soc.",
    volume = "112",
    pages = "195",
    year = "1952"
}

@article{Michel1972Accretion,
    author = "Michel, F. Curtis",
    title = "{Accretion of matter by condensed objects}",
    doi = "10.1007/BF00649949",
    journal = "Astrophys. Space Sci.",
    volume = "15",
    number = "1",
    pages = "153--160",
    year = "1972"
}

@article{AhmedEtAl2016Accretion,
 author = {Ahmed, Ayyesha K. and Azreg-A{\"\i}nou, Mustapha and Faizal, Mir and Jamil, Mubasher},
    title = "{Cyclic and heteroclinic flows near general static spherically symmetric black holes}",
    eprint = "1512.02065",
    archivePrefix = "arXiv",
    primaryClass = "gr-qc",
    doi = "10.1140/epjc/s10052-016-4112-y",
    journal = "Eur. Phys. J. C",
    volume = "76",
    number = "5",
    pages = "280",
    year = "2016"
}

@article{AzregAinou2017Accretion,
 author = {Azreg-A{\"\i}nou, Mustapha},
    title = "{Cyclic and heteroclinic flows near general static spherically symmetric black holes: Semi-cyclic flows -- Addendum and corrigendum}",
    eprint = "1605.06063",
    archivePrefix = "arXiv",
    primaryClass = "gr-qc",
    doi = "10.1140/epjc/s10052-017-4613-3",
    journal = "Eur. Phys. J. C",
    volume = "77",
    number = "1",
    pages = "36",
    year = "2017"
}

@article{Pantig:2024kqy, author = "Pantig, Reggie C.", title = "{On the analytic generalization of particle deflection in the weak field regime and shadow size in light of EHT constraints for Schwarzschild-like black hole solutions}", eprint = "2409.00476", archivePrefix = "arXiv", primaryClass = "gr-qc", doi = "10.1140/epjc/s10052-025-13766-1", journal = "Eur. Phys. J. C", volume = "85", number = "1", pages = "52", year = "2025" }

\end{document}